\documentclass[a4paper,fleqn]{cas-dc}

\usepackage[numbers]{natbib}

\usepackage{amssymb}

\usepackage{graphicx}
\usepackage{amsmath}
\usepackage{subfigure}
\usepackage{braket}
\usepackage{bbold}
\usepackage{yfonts}
\usepackage{placeins}
\usepackage{bm} 
\usepackage{nicefrac}
\usepackage{slashed}

\begin{document}
\let\WriteBookmarks\relax
\def\floatpagepagefraction{1}
\def\textpagefraction{.001}
\shorttitle{Rotational Brownian motion and heavy quark polarization in QCD medium}
\shortauthors{ }

\title [mode = title]{Rotational Brownian motion and heavy quark polarization in QCD medium}

\author[1]{Sourav Dey}[orcid=0009-0006-4272-755X]
\ead{sourav.dey@niser.ac.in}

\address[1]{School of Physical Sciences, National Institute of Science Education and Research, An OCC of Homi Bhabha National Institute, Jatni-752050, India}

\author[1]{Amaresh Jaiswal}[orcid=0000-0001-5692-9167]
\ead{a.jaiswal@niser.ac.in}
\cormark[1]


\begin{abstract}
We consider the rotational Brownian motion of heavy quark in QCD medium and provide analytical results for polarization of open heavy-flavor hadrons. We calculate expressions for vector and tensor polarization, corresponding to baryon spin polarization and vector meson spin alignment, respectively. Assuming that heavy quarks are initially fully spin polarized along the direction of magnetic field, we compare our results with recent experimental data from ALICE collaboration for $D^{*+}$ meson and provide predictions for spin polarization of open charm baryons. We propose that the transverse momentum dependence of heavy quark polarization may serve as a distinctive signature of the intense initial magnetic field generated in off-central relativistic heavy-ion collisions.
\end{abstract}


\maketitle
\section{Introduction}

The study of Brownian motion of heavy quarks through the quark-gluon plasma (QGP) formed in high-energy heavy-ion collisions has long been regarded as a crucial probe of its transport properties~\cite{Moore:2004tg, Gubser:2006bz, Das:2010tj, Akamatsu:2008ge, Banerjee:2011ra, Ding:2012iy, vanHees:2007me, PhysRevD.37.2484, dong2019heavy, PhysRevC.73.034913, das2015toward, PhysRevLett.100.192301, ALICE:2021rxa, STAR:2017kkh}. A key objective of phenomenological studies on heavy quark dynamics in QGP is to determine the drag and diffusion coefficients, which characterizes the interaction of heavy quarks with the surrounding medium. These studies typically focus on the translational Brownian motion of heavy quark where the drag and diffusion of the medium influences its linear momentum and spatial displacement. On the other hand, owing to the recent observation of hadron polarization in relativistic heavy-ion collisions~\cite{Liang:2004ph, Liang:2004xn, STAR:2017ckg, STAR:2018pps, STAR:2018fqv, Niida:2018hfw, STAR:2018gyt, ALICE:2019aid, STAR:2019erd, ALICE:2019onw, Singha:2020qns, Chen:2020pty, STAR:2020xbm, ALICE:2021pzu, STAR:2021beb, Mohanty:2021vbt, STAR:2022fan, STAR:2023eck}, it is of interest to consider rotational Brownian motion of heavy quarks as they propagate through the QGP.

Rotational Brownian motion refers to the random rotational motion (orientation and angular velocity) of a microscopic particle due to thermal fluctuations caused by collisions with surrounding medium particles. The problem of rotational Brownian motion was first considered by Debye, who applied Einstein's theory of Brownian motion to explain the anomalous dispersion observed at radio frequencies~\cite{Debye1929}. In his analysis, he examined a collection of molecules, each with a permanent dipole moment, constrained to rotate about an axis perpendicular to itself. Assuming negligible electrical interactions between molecules, he concluded that all molecules in the ensemble behave identically on average. Consequently, the problem reduces to studying the rotational Brownian motion of a single molecule in two dimensions, or equivalently, the motion of a dipole or rigid rotator under the influence of an external time-varying electric field. Since then, numerous studies have explored the theory of rotational Brownian motion and its applications to molecular systems and dielectrics~\cite{Perrin1934, Landau:1935qbc, Furry1957, Favro1960, Evanov1964, Hubbard1972, Valiev_1973, COFFEY1980, Coffey_2003,KuboSPINFIRST}. The study of heavy quark polarization within the framework of rotational Brownian motion is relatively unexplored~\cite{Liu:2024hii}.

Unlike light hadrons, which are usually produced near the freeze-out hypersurface, i.e., close to medium hadronization, heavy quarks are primarily generated in the initial hard scatterings of partons, making them a relatively clean probe of the early-stage properties of the hot deconfined medium. Moreover, in off-central relativistic heavy-ion collisions, strong transient magnetic fields are produced which decay very fast and are significant only during the early stages of the collision. These strong magnetic fields may induce spin polarization of heavy quarks along the magnetic field direction, which can manifest itself in hadron polarization. For instance, the induced spin polarization of heavy quarks can be observed in the polarization of open heavy-flavor hadrons, such as D-mesons or $\Lambda_c$ baryons. Therefore heavy quark polarization observables offers an unique probe into the initial strong magnetic field produced in relativistic heavy ion collisions.

In this Letter, we investigate the rotational Brownian motion of heavy quarks in a QCD medium and present analytical results for the polarization of open heavy-flavor hadrons. We start with the assumption that heavy quarks are fully spin polarized along the direction of the initial strong magnetic field and calculate the residual polarization after interaction with the QCD medium. Specifically, we derive expressions for vector and tensor polarization, which correspond to baryon spin polarization and vector meson spin alignment, respectively. We observe that heavy quarks with higher transverse momentum retain greater polarization due to their reduced interaction time with the medium. We propose this transverse momentum dependence of the heavy quark polarization as an unique signal for the initial strong magnetic field created in off-central relativistic heavy-ion collisions. We compare our findings with recent experimental data from the ALICE collaboration for the $D^{*+}$ meson~\cite{ALICE:2025cdf} and give predictions for the spin polarization of open charm baryons.

Throughout the text, three vectors are denoted in bold fonts and their scalar products are defined as $\bm{A}\cdot \bm{B}\equiv A_i B_i$. We use $\epsilon_{ijk}$ to denote the completely antisymmetric unit tensor in three indices (Levi-Civita symbol). We follow Einstein summation convention over repeated indices and use natural units $c=\hbar=k_B=1$

\section{Rotational Brownian motion}

For classical spins, the Langevin equation corresponds to the stochastic Landau–Lifshitz-Gilbert equation which governs the spin dynamics of a particle with magnetic moment in the particle's rest frame~\cite{Landau:1935qbc, Gilbert_2004, Nishino_2015, Meo_2023}. In terms of particle spin, the Landau–Lifshitz-Gilbert equation can be written as\footnote{Note that the motion of the relativistic classical spin of a particle is governed by the Thomas–Bargmann–Michel–Telegdi (Thomas–BMT) equation~\cite{Leader_2001}. In the particle’s rest frame, the spin dynamics described by the Thomas–BMT equation become independent of the particle's momentum and coincide with Eq.~\eqref{SpinLangevin} in the absence of stochastic and drag terms.}
\begin{equation}\label{SpinLangevin}
\frac{d\bm{s}}{d\tau} = \bm{s} \times \left[ \tilde{\bm{B}} + \boldsymbol{\xi}(\tau) \right] - \lambda\, \bm{s} \times \left( \bm{s} \times \tilde{\bm{B}} \right),
\end{equation}
where $\bm{s}$ represents the classical spin vector and $\tau$ is the proper time as measured in the particle rest frame. In the above equation, the effective field associated with the time dependent spin Hamiltonian $ \mathcal{H}(\bm{s})$ is given by $\tilde{\bm{B}} \equiv \gamma\bm{B} = -\frac{\partial \mathcal{H}}{\partial \bm{s}}$, where $\bm{B}$ is the external magnetic field as observed in the particle rest frame and $\gamma$ is the gyromagnetic ratio relating magnetic moment vector and spin vector, $\boldsymbol{\mu} = \gamma\, \bm{s}$. The term consisting of $\bm{s}\times\tilde{\bm{B}}$ represents precession dynamics of the system. The term involving the double vector product acts as a damping mechanism, guiding $\bm{s}$ toward potential minima while maintaining its magnitude. Here $\boldsymbol{\xi}(\tau)$ is the random torque experienced by the particle due to its interaction with the medium. The stochastic properties of the components of the fluctuating field $\boldsymbol{\xi}(\tau)$ follow the standard correlation rules, as described later in Eq.~\eqref{Correlation}. Lastly, the damping coefficient $\lambda$ quantifies the relative importance between relaxation and precession dynamics.

We note that Eq.~\eqref{SpinLangevin} is a special case of the generalized multivariate Langevin equation \cite{risken1996fokker,Balakrishnan}
\begin{equation}\label{mult_langevin}
    \frac{d y_i}{d\tau} = A_i(y, \tau) + C_{ik}(y, \tau) \, \xi_k(\tau),
\end{equation}
where $\xi_k(\tau)$ denotes the fluctuations. The statistical properties of the noise terms are \cite{PhysRevB.91.134411,PhysRevB.83.054432}
\begin{equation}\label{Correlation}
    \langle \xi_k(\tau) \rangle = 0, \quad
    \langle \xi_k(\tau_1)\, \xi_{l}(\tau_2) \rangle = 2\, D\, \delta_{kl} \,\delta(\tau_1 - \tau_2),
\end{equation}
which is a mathematical statement that the fluctuations are statistically independent, indicating that they constitute white noise. In the above equation, $D$ represents spin diffusion coefficient. Using the Kramers–Moyal expansion for Eqs.~\eqref{mult_langevin} and \eqref{Correlation}, one arrives at the Fokker–Planck equation \cite{risken1996fokker, Livi_Politi_2017}
\begin{align}
    \frac{\partial \mathcal{P}}{\partial \tau} = &\,
    - \frac{\partial}{\partial y_i} 
    \left[ A_i(y, \tau) + D \, C_{jk}(y, \tau) 
    \frac{\partial C_{ik}(y, \tau)}{\partial y_j} \right] \mathcal{P} \nonumber\\
    &\,+ D \, \frac{\partial^2}{\partial y_i \partial y_j} 
    \left[ C_{ik}(y, \tau) C_{jk}(y, \tau) \mathcal{P} \right],
    \label{eq:Genfokker_planck}
\end{align}
where, $\mathcal{P} \equiv\mathcal{P}(y,\tau|y_0,\tau_0)$ is the transition probability from the state $(y_0,\tau_0)$ to $(y,\tau)$. It is important to note that this equation is entirely determined by the coefficients of the Langevin equation. 

The stochastic Landau–Lifshitz-Gilbert equation, Eq.~\eqref{SpinLangevin}, can be recast in the form of general multivariate Langevin equation, \eqref{eq:Genfokker_planck}, by identifying $s_i=y_i$ and 
\begin{align}
    A_i &=  \epsilon_{ijk} s_j \tilde{B}_k + \lambda\, (s^2 \delta_{ik} - s_i s_k) \tilde{B}_k, \label{eq:A_coefficient} \\
    C_{ik} &= \epsilon_{ijk} s_j. \label{eq:B_coefficient}
\end{align}
In order to arrive at the above relations, we have used the identities
\begin{align}
\frac{\partial C_{ik}}{\partial s_j} =&\, \epsilon_{ijk}, \quad  C_{jk} \, \frac{\partial C_{ik}}{\partial s_j} = -2\,s_i,  
\label{eq:diffusion_derivative}\\
C_{ik} C_{jk} =&\, s^2 \delta_{ij} - s_i s_j. \label{CikCjk}
\end{align}
This allows us to formulate the Fokker–Planck equation corresponding to the stochastic Landau–Lifshitz-Gilbert equation as~\cite{PhysRevB.91.134411, PhysRevB.83.054432, PhysRevB.58.14937}
\begin{align}
    \frac{\partial \mathcal{P}}{\partial \tau} = &\,
    -\! \frac{\partial}{\partial s_i} 
    \left[
        \epsilon_{ijk}\, s_j \tilde{B}_k + \lambda
        (s^2 \delta_{ik} - s_i\, s_k) \tilde{B}_k \!
        - 2D s_i \right] \mathcal{P} \nonumber\\
    &\, + D \, \frac{\partial^2}{\partial s_i\, \partial s_j} 
    \left[ s^2 \delta_{ij} - s_i s_j \right] \mathcal{{P}}, \label{delPdelt}
\end{align}
where $\mathcal{P}\equiv \mathcal{P}(\bm{s},\tau)$ is the probability distribution for spin orientation along $\textbf{s}$ at time $\tau$. We note that corrections to the above equation arising from the inclusion of a fluctuating field in the damping term of Eq.~\eqref{SpinLangevin} appear at order $D\lambda^2$ (see Appendix~\ref{app:fluc_th}) and are therefore neglected in the present analysis.

in presence of a fluctuating field on the second (thermal) term in the right hand side of Eq.~\eqref{SpinLangevin}

Assuming that the field $\tilde{\bm{B}}$ is external and independent of particle spin $\bm{s}$, Eq.~\eqref{delPdelt} reduce to a Smoluchowski-like equation~\cite{LEcturenotes1}
\begin{equation}
    \frac{\partial \mathcal{P}}{\partial \tau} =
    D\, \frac{\partial}{\partial \bm{s}} \cdot 
    \left[
        \bm{s} \times 
        \left(
            \bm{s} \times 
            \left( \frac{\lambda}{D} \tilde{\bm{B}} - \frac{\partial}{\partial \bm{s}} \right)
        \right)
    \right] \mathcal{P},
    \label{eq:smoluchowski}
\end{equation}
which can be seen as the rotational counterpart of the Klein– Kramers equation~\cite{RevModPhys.15.1}. Here, our goal is to determine the probability of a spin-polarized particle having an instantaneous orientation in the direction $(\theta,\phi)$. This corresponds to considering a sphere in spin-space of fixed radius $s$, i.e., $\bm{s}=(s,\theta,\phi)$, where each point on the sphere represents a different spin orientation for the particle \cite{PhysRevE.76.051104, PhysRevA.11.280}. Without loss of generality, we consider the direction of the external field $\tilde{\bm{B}}$ along $z$-direction. Assuming an axially symmetric Hamiltonian, the equation simplifies to \cite{Debye1929, PhysRevE.76.051104, PhysRev.130.1677}
\begin{equation} \label{eq:spherical_fokker_planck}
    \tau_s \frac{\partial \mathcal{P}}{\partial \tau} =
    \frac{1}{\sin \theta} \frac{\partial}{\partial \theta} 
    \left[
        \sin \theta 
        \left(\frac{\lambda}{D}
            \frac{\partial \mathcal{H}}{\partial \theta} \mathcal{P} +  \frac{\partial \mathcal{P}}{\partial \theta}
        \right)
    \right],
\end{equation}
where, $\tau_s\equiv 1/D$ represents spin relaxation time\footnote{Similar to linear Brownian motion, the spin relaxation time should be related to dissipation in spin hydrodynamics through an Einstein-Stokes-like relation~\cite{BitaghsirFadafan:2008adl}. Nevertheless, an estimate for $\tau_s$ can be obtained from energy/length scales relevant to the system~\cite{Hongo:2022izs, Hidaka:2023oze}.} and $\mathcal{H}=-\boldsymbol{\mu}\cdot\bm{B}=-\bm{s}\cdot\tilde{\bm{B}}$. For the case of a spatially homogeneous and time varying magnetic field $\textbf{B}$, the above equation reduces to \cite{KuboSPINFIRST, PhysRevA.11.280}
\begin{equation}\label{FokPM}
    \tau_s \frac{\partial \mathcal{P}}{\partial \tau} = 
    \frac{1}{\sin \theta} \frac{\partial}{\partial \theta} \Bigg[ \sin \theta \Bigg(\frac{\partial}{\partial \theta} + \frac{\lambda}{D}\mu B(\tau) \sin \theta\, \Bigg) \Bigg]\mathcal{P}\, , 
\end{equation}
where $\mu$ is the magnitude of the magnetic moment of the particle. In the following, we obtain the solution of the above equation, which represents the time evolution of the diffusion of spin polarization.

\section{Heavy quark polarization}

The Fokker-Planck equation given in Eq.~\eqref{FokPM} can be written in shorthand notation as
\begin{align}\label{fokker_planck}
    \tau_s\, \partial_{\tau}\mathcal{P}(\theta,\tau) = \mathcal{L}_{\theta}(\tau) \, \mathcal{P}(\theta,\tau)
\end{align}
Here, $\partial_\tau\equiv\frac{\partial}{\partial \tau}$ and $\mathcal{L}_\theta (\tau)$ is a time-dependent differential operator with respect to $\theta$. Considering the time evolution in operator form,
\begin{align}\label{fokpl_ope}
    \tau_s\,\partial_{\tau}\ket{\mathcal{P},\tau}=\hat{\mathcal{L}}(\tau)\ket{P,\tau}.
\end{align}
It is easy to see that the generic solution has the structure
\begin{align}
    \ket{\mathcal{P},\tau}=\text{exp}\left[ \frac{1}{\tau_s}\int_{0}^{\tau}d\tau^\prime\, \hat{\mathcal{L}}(\tau^\prime)\right] \ket{\mathcal{P},0} .
\end{align}
We note that the differential operator $\hat{\mathcal{L}}$ can be separated into a time-independent as well as time-dependent parts, i.e., $\hat{\mathcal{L}}^{0}$ and $\hat{\mathcal{L}}^\prime(\tau)$ respectively. Considering the time dependence of the magnetic field to be of the form $B(\tau)=B_{0}\,\phi(\tau)$, we obtain
\begin{equation}
\hat{\mathcal{L}}(\tau)=\hat{\mathcal{L}}^{0}+\alpha \, \hat{\mathcal{L}}^\prime(\tau),
\end{equation}
where $\alpha\equiv \mu B_0\lambda/D$. Further, assuming $\alpha$ to be a small parameter (which is later shown to be valid for the case of heavy quarks in QGP), we invoke the Dyson series expansion in powers of $\alpha$ to obtain the solution \cite{risken1996fokker,PhysRevResearch.3.043172}
\begin{eqnarray}
&\!\!\!\!\!\!\mathcal{P}\,(\theta,\tau;\theta_{0},0) \equiv \bra{\theta}\ket{\mathcal{P},\tau} =\mathcal{P}^{0}(\theta,\tau;\theta_{0},0) +\alpha\!\!\int_\Omega\!\int_{0}^{t}\!\!\frac{d\tau^\prime}{\tau_s}d\Omega^\prime \nonumber\\
&\mathcal{G}^{0}_{\tau-\tau^{\prime}}(\theta,\theta^\prime)\mathcal{L}^\prime_{\theta^\prime}(\tau^\prime)\mathcal{P}^{0}(\theta^\prime\!,\tau^\prime;\theta_{0},0) \!+\! \mathcal{O}(\alpha^2) \,, \label{SolDyson} 
\end{eqnarray}
where, $d\Omega^\prime \equiv 2\pi \sin\theta^\prime\,d\theta^\prime$. In the above equation, $\mathcal{P}^{0}(\theta,\tau;\theta^\prime,0)$ and $\mathcal{G}_{\tau}^{0}(\theta,\theta^\prime)$ are the solution and the corresponding Green function of Eq.~\eqref{fokker_planck}, respectively, for $\alpha=0$.

In the absence of the magnetic field, the operator $\hat{\mathcal{L}}$ in Eq.~\eqref{fokpl_ope} consists of only the time-independent part $\hat{\mathcal{L}}^{0}$, which is given by
\begin{equation}
    \hat{\mathcal{L}}^{0} = \frac{1}{\sin \theta} \frac{\partial}{\partial \theta} \bigg( \sin \theta \frac{\partial}{\partial \theta} \bigg).
\end{equation}
In this case, solution of Eq.\,\eqref{fokker_planck} can be expressed as 
\begin{align}
    \mathcal{P}(\theta,\tau ;\theta^\prime,0)=\int d\Omega^\prime \, \mathcal{G}_{\tau}(\theta,\theta^\prime) \, \mathcal{P}(\theta^\prime,0)\,,
\end{align}
where,
\begin{align}
    \mathcal{G}_{\tau}(\theta,\theta^\prime)=\sum_{n=0}^{\infty}\frac{(2n+1)}{4\pi}\exp\left[-n(n+1)\frac{\tau}{\tau_s}\right]\nonumber\\
    \times P_{n}(\cos\theta) \, P_{n}(\cos\theta^\prime).
\end{align}
Here, $P_{n}(x)$ are Legendre polynomials in $x$ of degree $n$. Assuming that all heavy quarks are initially spin polarized along the magnetic field direction $\theta=\theta_{0}$, i.e., for the initial condition $\mathcal{P}(\theta,0) = \frac{1}{2\pi}\delta(\cos\theta-\cos\theta_{0})$, we obtain 
\begin{eqnarray}
    \displaystyle{ \mathcal{P}(\theta,\tau ;\theta_{0},0) = \sum_{n=0}^{\infty}\frac{(2n+1)}{4\pi}\exp\left[-n(n+1)\frac{\tau}{\tau_s}\right] } \nonumber\\
    \displaystyle{\times P_{n}(\cos\theta)P_{n}(\cos\theta_{0})}.
\end{eqnarray}
The above solution in the absence of the magnetic field, as obtained in the above equation, forms the basis for the perturbative treatment of the time-dependent magnetic field. Note that the primary effect of the strong initial magnetic field is incorporated through the initial condition, where we assume all heavy quark spins are aligned with the magnetic field direction.

Next, we consider a time-dependent magnetic field characterized by the perturbative parameter $\alpha$, where the corresponding time-dependent operator is given by
\begin{align}
    \mathcal{L}^\prime(\tau)=\frac{\phi(\tau)}{\sin\theta}\frac{\partial}{\partial\theta}\,\sin^{2}\theta\,.
\end{align}
Using the expression given in Eq.~\eqref{SolDyson}, one can directly calculate the expectation value of $\cos\theta$ and $\cos^2\theta$, which quantifies vector and tensor polarizations, respectively~\cite{Leader_2001}. Assuming that the polarization of open heavy-flavor hadrons will be dominated by the heavy quark polarization,  the vector polarization corresponds to baryon spin polarization whereas the tensor polarization is relevant for vector meson spin alignment. This picture is consistent with the mechanism of polarized heavy quark hadronization via fragmentation. Using Eq.~\eqref{SolDyson}, we obtain
\begin{eqnarray}
    \displaystyle{ \langle \cos\theta \rangle = \cos\theta_{0} \, e^{- 2\tau/\tau_s} - \frac{2\alpha}{3\tau_s }\, e^{-2\tau/\tau_s}\int_{0}^{\tau} \! d\tau^\prime \, \phi(\tau^\prime)} \nonumber\\
    \displaystyle{ \Big[e^{2\tau^\prime/\tau_s} - P_{2}(\cos\theta_{0}) e^{-4\tau^\prime/\tau_s}\Big]}\,,\label{timPol}
\end{eqnarray}
and
\begin{eqnarray}
     \displaystyle{\langle \cos^2\theta \rangle = \frac{1}{3} + \frac{2}{3} P_2( \cos\theta_{0}) e^{- 6\tau/\tau_s} + \frac{2\alpha}{5\tau_s}\! \int_{0}^{\tau} \! d\tau^\prime \phi(\tau^\prime)} \nonumber\\
     \displaystyle{\Big[P_1( \cos\theta_{0})e^{(4\tau^\prime-6\tau)/\tau_s} - P_3( \cos\theta_{0})e^{-6(\tau^\prime+\tau)/\tau_s} \Big]}\,.\label{timPol2}
\end{eqnarray}
We aim to calculate the spin polarization of heavy quarks induced by the initial strong magnetic field. This field causes the heavy quark spins to align either parallel or anti-parallel to the magnetic field, depending on the quark's charge. Consequently, the relevant values of $\theta_0$ in this scenario are $0$ and $\pi$, for which $P_2(\cos\theta_0) = 1$.

The magnetic field produced in relativistic heavy-ion collisions is short lived and decays very fast due to the spectators receding away from the collision zone at relativistic velocities. In this case, one can consider an exponentially decaying profile of the magnetic field, $\phi(\tau)=e^{-\tau/\tau_B}$, where $\tau_B$ represents typical timescale over which the magnetic field survives. Using this form of the time-dependence, we perform the integral in Eqs.~\eqref{timPol} and Eqs.~\eqref{timPol2} to obtain
\begin{eqnarray}
    &\!\!\!\!\!\!\!\!\!\!\!\!\!\!\!\displaystyle{\langle \cos\theta \rangle = \cos\theta_{0} \, e^{- 2\tau/\tau_s} -\frac{2\alpha\tau_{B}}{3} \, e^{- 2\tau/\tau_s}} \nonumber\\
     &\!\!\!\!\!\!\!\!\!\!\displaystyle{\left[\frac{1-\exp\left(-\frac{(\tau_s-2\tau_{B})\tau}{\tau_s\tau_{B}}\right)}{\tau_s-2\tau_{B}} -\frac{1-\exp\left(-\frac{(4\tau_{B}+\tau_s)\tau}{\tau_s\tau_{B}}\right)}{4\tau_{B}+\tau_s} \right]}\, , \label{av_pol}
\end{eqnarray}
and
\begin{eqnarray}
    &\!\!\!\!\!\!\!\!\!\!\!\!\!\!\!\!\!\!\!\!\displaystyle{\langle \cos^2\theta \rangle = \frac{1}{3} + \frac{2}{3} \, e^{-6\tau/\tau_s} + \frac{2\alpha\tau_{B}}{5} e^{-6\tau/\tau_s}} \nonumber\\ 
    &\!\!\!\!\!\!\!\!\!\!\!\!\!\!\!\displaystyle{\times \Bigg[\frac{1}{(\tau_s-4\tau_B)} 
    \bigg( 1-\exp\Big[-\frac{(\tau_s-4\tau_B)\tau}{\tau_s\tau_B} \Big]
    \bigg)}  \nonumber\\
    &\!\!\!\!\!\!\!\!\!\!\!\!\!\!\!\displaystyle{-\frac{1}{(\tau_s+6\tau_B)} \bigg( 1-\exp\Big[-\frac{(\tau_s+6\tau_B)\tau}{\tau_s\tau_B} \Big] \bigg) \Bigg]\cos\theta_{0}}\, , \label{av_pol2}
\end{eqnarray}
where we have used $P_{2n}(\cos\theta_{0})=1$ and $P_{2n+1}(\cos\theta_{0})=\cos\theta_{0}$ keeping in mind that $\theta_{0}=0,\pi$.

Next we consider heavy quarks with longitudinal rapidity $y$ having transverse velocity $v_{T}\equiv p_T/E$, where $p_T$ is the transverse momentum, $E=m_T\cosh y$ is the relativistic particle energy and $m_T=\sqrt{p_T^2+m_Q^2}$ is defined as the transverse mass with $m_Q$ being the heavy quark mass. Keeping in mind that the entire formulation in the present work has been performed in the heavy quark rest frame, the magnetic field along $y$-direction, created due to spectators in the lab frame, has to be Lorentz transformed to the heavy quark rest frame. The Lorentz transformation of the initial electromagnetic field in the heavy quark rest frame is given by~\cite{Herbert:1997}
\begin{equation}\label{Lorentz_trans_B}
\bm{B} = \gamma_{v} \left( \bm{B_{\rm Lab}} - \bm{E_{\rm Lab}} \times \bm{v} \right) + \left( 1 - \gamma_{v} \right) \left( \frac{\bm{B_{\rm Lab}}\cdot\bm{v}}{|\bm{v}|^2} \right) \bm{v},
\end{equation}
where $\gamma_{v}\equiv E/m_Q$ is the Lorentz gamma factor, $\bm{v}$ is the velocity of the heavy quark and, $\bm{E_{\rm Lab}}$ and $\bm{B_{\rm Lab}}$ are laboratory-frame electric and magnetic fields, respectively. From the above equation, we infer that for a given $p_T$, the magnetic field component along the $y$-direction in the heavy quark's rest frame increases with rapidity. Therefore, the signal of spin polarization induced by the initial magnetic field is expected to be more prominent for heavy quarks with finite rapidity because it experiences larger magnetic field in its rest frame. 

Owing to relativistic effects, the direction of the magnetic field in the heavy-quark rest frame generally differs from that in the laboratory frame, as indicated by Eq.~\eqref{Lorentz_trans_B}. Consequently, heavy quarks are assumed to be fully polarized along the magnetic-field direction in their rest frame. The initial polarization direction can therefore be characterized by an angle $\theta_{0}$ with respect to the fixed axis defined by $\bm{B_{\rm Lab}}$. This angle depends on the initial velocity of the heavy quark as well as on the orientations of $\bm{B_{\rm Lab}}$ and $\bm{E_{\rm Lab}}$. Upon averaging over an ensemble of heavy quarks, the polarization signal is expected to remain sensitive only to the direction of $\bm{B_{\rm Lab}}$. The dependence on $\bm{E_{\rm Lab}}$ cancels upon averaging due to the random orientations of the initial heavy-quark velocities (see Appendix~\ref{app:averaging}). This behavior is consistent with the physical picture that the magnitude of the observed polarization reflects the extent to which the initial spin orientation is randomized through interactions with the medium. Such randomization is reduced at higher transverse momentum $p_T$ for a given rapidity $y$, leading to a correspondingly larger polarization signal.

In order to validate the perturbative treatment of the time-dependent magnetic field, we estimate the maximum value of $\alpha$. The initial magnetic field in the lab frame is along the $y$-direction and typically reduces to $eB_{\rm Lab}\lesssim 0.1 m_{\pi}^{2}$ by the time the QGP is formed. The perturbation parameter $\alpha$ can be estimated as\footnote{In absence of Einstein-Stokes-like relation between drag and diffusion coefficients for spin evolution, we assume a similar relation based on dimensional arguments, i.e., $D=\lambda T$.}
\begin{equation}
    \alpha \equiv \frac{\mu\, |\mathbf{B}|\lambda}{D}  \sim \frac{{\rm g}\, s\, q\, \gamma_v |\mathbf{B}_{\rm Lab}|}{2\, m_QT} = \frac{(f\,e) \gamma_v |\mathbf{B}_{\rm Lab}|}{2\, m_{Q}T},
\end{equation}
where for heavy quarks, we have considered the magnetic moment $\mu={\rm g}\,s\,q/(2m_Q)$, the g-factor to be ${\rm g} \approx 2$, the spin $s=\hbar/2$ and the charge $q=f\,e$, where $f$ is $2/3$ and $-1/3$ for charm and bottom quarks, respectively. Using these values and assuming an average QGP temperature of $300$~MeV we find that for heavy quarks with momentum $50$~GeV, the coefficient $\alpha$ evaluates to $0.034$ for charm quarks and $-0.007$ for bottom quarks. These values justify treating $\alpha$ as a perturbation parameter, validating the perturbative approach followed earlier. 

At this juncture, we note that the initial magnetic field diminishes considerably by the time the QGP is formed~\cite{Tuchin:2013ie, Huang:2022qdn, PhysRevC.105.054907, Huang:2017tsq, McLerran:2013hla}. Moreover, the spin relaxation time is predicted to be significantly larger than the QGP formation time~\cite{Hongo:2022izs, Hidaka:2023oze, Kapusta:2019sad}. Therefore, in the limit $\tau_s\gg\tau_B$, it is a reasonable approximation to neglect the $\alpha$-dependent terms in Eq.~\eqref{av_pol} and \eqref{av_pol2}, leading to
\begin{equation}\label{vec_ten_pol}
    \langle \cos\theta \rangle = \cos\theta_{0} \, e^{- 2\tau/\tau_s} , \quad
    \langle \cos^2\theta \rangle = \frac{1}{3} + \frac{2}{3} \, e^{-6\tau/\tau_s} .
\end{equation}
In the above equations, $\tau$ denotes the total duration for which the heavy quark undergoes Brownian motion within the QGP. Therefore, we conclude that longer the heavy quarks remain in the QGP, the smaller their spin polarization becomes. Further, we observe that $\tau_s = 1/D$ is the only free parameter in Eq.~\eqref{vec_ten_pol}, and it represents the interaction of the heavy quark with the medium. It is important to note that the spin relaxation time $\tau_s$ depends on the QGP properties like temperature and has been estimated earlier in the context of strange quarks~\cite{Hongo:2022izs, Hidaka:2023oze, Kapusta:2019sad}. On the other hand, since heavy-quark energy loss is not included in the present analysis, the interpretation of $\tau_s$ as a spin relaxation time should be regarded only as an order-of-magnitude estimate.

From Eq.~\eqref{vec_ten_pol}, it is also clear that the vector polarization, $\langle \cos\theta \rangle$, is sensitive to charge of the heavy quarks, as particles with opposite charges will be polarized in opposite directions ($\cos\theta_0=\pm1$). On the other hand, we do not observe such features for the case of tensor polarization $\langle \cos^2\theta \rangle$. Therefore, we conclude that the polarization of open heavy baryons (such as $\Lambda_c$) and anti-baryons will have opposite signs, whereas open heavy mesons and anti-mesons will exhibit polarization of the same sign. Moreover, the polarization of heavy quarkonia induced by the initial magnetic field is expected to be small due to the opposite polarization of heavy quarks and anti-quarks in magnetic field~\cite{ALICE:2020iev}.

\begin{figure}
    \centering
    \includegraphics[width=\linewidth]{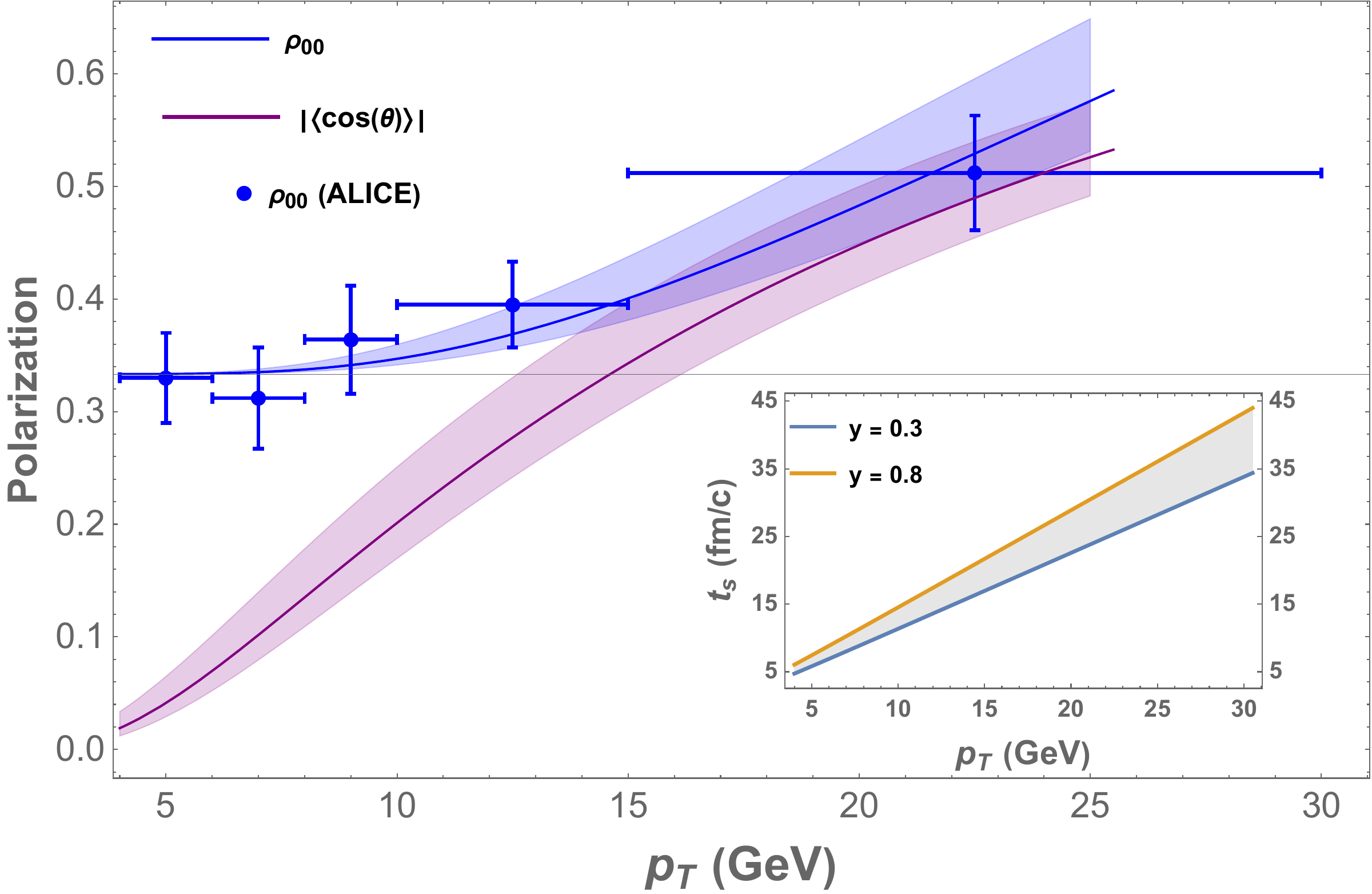}
    \caption{The ALICE measurements of the spin alignment parameter $\rho_{00}$ for the $D^{*+}$ meson are presented as a function of transverse momentum~\cite{ALICE:2025cdf}. Our best-fit result for $\rho_{00}$, obtained using Eq.~\eqref{vec_ten_pol}, is shown. We also show predictions for polarization of open-charmed baryons, expressed as $|\langle \cos\theta \rangle|$. The inset illustrates the spin-relaxation time in the lab frame as a function of $p_T$. The shaded regions indicates the rapidity window of the experimental measurement, $0.3 < |y| < 0.8$.}
    \label{Fig}
\end{figure}
 
We denote the average path length traversed by a heavy quark in the fireball as $L$, which typically depends on the colliding system as well as centrality of the collision. Here, we assume $L\simeq 10$~fm as a typical order of magnitude. Considering Lorentz contraction of $L$ in the heavy quark rest frame, the duration for which it undergoes Brownian motion within the QGP is given by $\tau=L\, m_{Q}/|{\bm p}|$, where $|{\bm p}|=\sqrt{p_{T}^{2}\cosh^{2}y+m_{Q}^{2}\,\sinh^{2}y}$. Therefore the final expression for polarization depends on the transverse momentum of the heavy quarks and its rapidity. From the expressions for vector and tensor polarizations given in Eq.~\eqref{vec_ten_pol}, we see that heavy quark polarization increases with increase in transverse momentum. In order to compare with experimental results, we express $\rho_{00}=1/3 + (5/2)\left[\langle \cos^2\theta \rangle -1/3\right]$.

In Fig.~\ref{Fig}, we show the best fit result for $\rho_{00}$ measured by the ALICE collaboration for the $D^{*+}$ meson (blue dots) for $\sqrt{s_{\rm NN}}=5.02$~TeV Pb$-$Pb collisions at $30-50\%$ centrality and in the rapidity window $0.3 < |y| < 0.8$~\cite{ALICE:2025cdf}. We use Eq.~\eqref{vec_ten_pol} for the fit function (blue line) with $|y|=0.55$ and find that the best fit result is obtained for $\tau_s=1.31$~fm with a reduced $\chi^2 = 0.27$. In the inset, we show the spin-relaxation time in the lab frame, $t_s=\gamma_v\tau_s$, which increases with $p_T$. In Fig.~\ref{Fig}, we also show our prediction for open-charmed baryons (magenta line) using the fitted $\tau_s$. The increasing trend of $\rho_{00}$ and $|\langle\cos\theta\rangle|$ with $p_T$ can be attributed to the fact that faster heavy quarks exit the fireball more quickly, reducing their interaction time with the QCD medium and allowing them to retain a higher degree of spin polarization.

It is important to note that if the spin polarization of heavy quarks were solely due to their interactions with a polarized medium, the observed trend with respect to $p_T$ would be opposite to that shown in Fig.~\ref{Fig}. This is because heavier quarks traveling at lower velocities spend more time in the medium, allowing greater equilibration of their spin with the medium's polarization. In contrast, faster quarks interact for shorter durations, leading to less spin alignment. As a result, such a mechanism would produce a decreasing spin polarization with increasing $p_T$. Therefore, the observed increase in heavy quark spin polarization with $p_T$ cannot be attributed to polarization due to interaction with a polarized medium alone.
%

\section{Summary and outlook}

In this Letter, we investigated the rotational Brownian motion of heavy quarks in a QCD medium and presented results for the polarization of open heavy-flavor hadrons. We started with the assumption that heavy quarks are fully spin-polarized along the initial magnetic field direction and calculated the residual polarization after their interaction with the QCD medium. Specifically, we derived expressions for vector and tensor polarization, corresponding to baryon spin polarization and vector meson spin alignment, respectively. Our analysis revealed that heavy quarks with higher transverse momentum retain greater polarization due to their shorter interaction time with the medium. As a result, we proposed the increasing polarization as a function of transverse momentum as a distinctive signature of the strong initial magnetic field generated in off-central relativistic heavy-ion collisions. We compared our results with recent experimental data from the ALICE collaboration for the $D^{*+}$ meson and provided predictions for the spin polarization of open-charmed baryons.

Our work presents the first formulation of stochastic dynamics of spin polarization in the context of heavy quarks in relativistic nucleus–nucleus collisions and offers a theoretical explanation for experimental observations. This framework holds enormous potential for advancing our understanding of heavy quark and quarkonium polarization in relativistic nucleus–nucleus collisions. Furthermore, it opens a novel research direction involving Langevin and Fokker–Planck equations in angular momentum space, which we believe has the potential for significant future developments.

One of the shortcomings of the present work is that the fireball was assumed to be static with a constant average temperature. Accordingly, the parameter values and predictions obtained within the present model should be interpreted at a qualitative level, providing only order-of-magnitude estimates. Looking forward, it will be quantitatively relevant to consider realistic hydrodynamic evolution of the fireball. For instance, incorporating the change in heavy quark momentum due to collisional and radiative energy loss, within realistic numerical simulations, is essential for a precise determination of $\tau_s$. Moreover, heavy quark spin relaxation time, $\tau_s$, in presence of external magnetic field need to be calculated from field theoretic approach. Further, it will be interesting to derive an Einstein-Stokes-like relation between the spin diffusion coefficient and the dissipative parameters in spin hydrodynamics. Furthermore, it is desirable to obtain the Fokker-Planck equation for spin evolution from a kinetic theory framework with spin~\cite{Weickgenannt:2024ibf}. We leave these questions for future work.

\bigskip
{\bf Acknowledgements:} The authors express their gratitude to Sourav Kundu for valuable discussions which initiated this work. The authors gratefully acknowledge Department of Atomic Energy (DAE), India for financial support.

\appendix
\section{Fluctuations in the thermal term}
\label{app:fluc_th}

\medskip

In the following, we re-derive the Fokker-Planck equation in presence of a fluctuating field on the second (thermal) term  \cite{Liu:2024hii, Nishino_2015} in the right hand side of Eq.~\eqref{SpinLangevin}. The Landau–Lifshitz–Gilbert equation becomes
\begin{equation}\label{SpinLangevinA}
\frac{d\bm{s}}{d\tau} = \bm{s} \times \left[ \tilde{\bm{B}} + \boldsymbol{\xi}(\tau) \right] - \lambda\, \bm{s} \times \left( \bm{s} \times \left[\tilde{\bm{B}}+\boldsymbol{\xi}(\tau) \right]\right),
\end{equation}
Comparing with the general form the Langevin equation,
\begin{equation}\label{mult_langevinA}
    \frac{d y_i}{d\tau} = A_i(y, \tau) + C_{ik}(y, \tau) \, \xi_k(\tau),
\end{equation}
and its corresponding Fokker-Planck equation,
\begin{align}
    \frac{\partial \mathcal{P}}{\partial \tau} = &\,
    - \frac{\partial}{\partial y_i} 
    \left[ A_i(y, \tau) + D \, C_{jk}(y, \tau) 
    \frac{\partial C_{ik}(y, \tau)}{\partial y_j} \right] \mathcal{P} \nonumber\\
    &\,+ D \, \frac{\partial^2}{\partial y_i \partial y_j} 
    \left[ C_{ik}(y, \tau) C_{jk}(y, \tau) \mathcal{P} \right],
    \label{eq:Genfokker_planckA}
\end{align}
the stochastic Landau–Lifshitz-Gilbert equation, Eq.~\eqref{SpinLangevin}, can be recast in the form of general multivariate Langevin equation, \eqref{eq:Genfokker_planck}, by identifying $s_i=y_i$ and 
\begin{align}
    A_i &=  \epsilon_{ijk} s_j \tilde{B}_k + \lambda\, (s^2 \delta_{ik} - s_i s_k) \tilde{B}_k, \label{eq:A_coefficientA} \\
    C_{ik} &= \epsilon_{ijk} s_j- \lambda(s_{i}s_{k}-s^{2}\delta_{ik}). \label{eq:B_coefficientA}
\end{align}
In order to arrive at the above relations, we have used the identities
\begin{align}
\frac{\partial C_{ik}}{\partial s_j} =&\, \epsilon_{ijk}-\lambda(\delta_{ij}s_{k}+s_{i}\delta_{jk}-2\delta_{ik}s_{j}), \\  
C_{jk} \, \frac{\partial C_{ik}}{\partial s_j} =&\, -2\,s_i - 2\lambda^{2} s^{2}s_{i},  
\label{eq:diffusion_derivativeA}\\
C_{ik} C_{jk} =&\, s^2 \delta_{ij} - s_i s_j + \lambda^{2}s^{2}(s_{i}s_{j}-\delta_{ij}s^{2}). \label{CikCjkA}
\end{align}
This allows us to formulate the Fokker–Planck equation corresponding to the stochastic Landau–Lifshitz-Gilbert equation as
\begin{eqnarray}
    \frac{\partial \mathcal{P}}{\partial \tau} &\!\!\!\!\!\!=\!\!\!\!\!\!&
    - \frac{\partial}{\partial s_i}\!\! 
    \left[
        \epsilon_{ijk} s_j \tilde{B}_k \!+\! \lambda
        (s^2 \delta_{ik} \!- s_i s_k) \tilde{B}_k \!
        - 2D s_i \!- 2D \lambda^{2} s^{2}s_{i} \right]\! \mathcal{P} \nonumber\\
    &\!\!\!\!\!\!& + D \frac{\partial^2}{\partial s_i\, \partial s_j}\! 
    \left[ (s^2 \delta_{ij} - s_i s_j) + \lambda^{2}s^{2}(s^{2}\delta_{ij}-s_{i}s_{j}) \right] \mathcal{{P}}, \label{delPdeltAA}
\end{eqnarray}
where $\mathcal{P}\equiv \mathcal{P}(\bm{s},\tau)$ is the probability distribution for spin orientation along $\textbf{s}$ at time $\tau$. In the above equation, we see that corrections to the Fokker-Planck equation, due to inclusion of fluctuating field in the second (thermal) term, appears at $D\lambda^2$ order. These corrections contribute at $\mathcal{O}(\tau_s^{-3})$ compared to the leading order result obtained in the main text, which appears at $\mathcal{O}(\tau_s^{-1})$. Since the heavy quark spin-relaxation time is expected to be large, we ignore $\mathcal{O}(\tau_s^{-3})$ terms in the present work.


\section{Averaging over heavy-quark velocity}
\label{app:averaging}

\medskip

The dominant magnetic field arising from spectator contributions is along the $y$-direction, while the electric field $\bm{E_{\rm Lab}}$ is along the $x$-direction in the lab frame~\cite{Gursoy:2014aka, Tuchin:2012mf}. If we assume that initial heavy quarks are produced with velocities distributed in all directions, then by averaging over the heavy quark distribution, one can show that
\begin{align}
   \big\langle \mathbf{B}\big\rangle&= \int \frac{d^{2}\Omega_v}{4\pi} f_{Q}(\theta_v,\phi_v) \Big[\gamma_{v} \left( \bm{B_{\rm Lab}} - \bm{E_{\rm Lab}} \times \bm{v} \right) \nonumber\\
   &+ \left( 1 - \gamma_{v} \right) \left( \frac{\bm{B_{\rm Lab}}\cdot\bm{v}}{|\bm{v}|^2} \right) \bm{v}\Big]\,.
\end{align}
Here, $\theta_v$ and $\phi_v$ are defined as the angles specifying the direction of the initial heavy quark velocity with respect to the fixed axis along $\bm{B_{\rm Lab}}$. Due to the geometry of the heavy-ion collision, the electric field $\bm{E_{\rm Lab}}$ from the spectator is oriented perpendicular to $\bm{B_{\rm Lab}}$. Assuming a uniform angular distribution of the initial heavy quark velocity, i.e., $f_{Q}(\theta_v,\phi_v)=\text{const.}$, the average over the magnetic field $\bm{B_{\rm Lab}}$ can be written as
\begin{align}
     \big\langle \mathbf{B}\big\rangle=\frac{1}{3}(1+2\gamma_{v} )\bm{B_{\rm Lab}}\,.\label{avB}
\end{align}
Hence, ensemble average over all possible heavy-quark polarizations shows independence from the electric field direction but reveals a clear signature along the magnetic field in lab frame. This is consistent with experimental observations, which indicate that after averaging over final particle polarizations, only the initial polarization component along the magnetic field survives. 
Furthermore, the ensemble average evaluated in Eq.~\eqref{avB} indicates that the contribution from the initial polarization is enhanced at large transverse momentum $p_{T}$ and rapidity $y$. This behavior follows directly from the Lorentz factor $\gamma_{v}=\frac{m_{T}}{m_{Q}}\cosh(y)$, which amplifies the polarization signal in this kinematic regime.

 
\printcredits

\bibliographystyle{model6-num-names}

\bibliography{cas-refs}

\begin{thebibliography}{75}
\providecommand{\natexlab}[1]{#1}
\providecommand{\url}[1]{\texttt{#1}}
\providecommand{\href}[2]{#2}
\providecommand{\path}[1]{#1}
\providecommand{\DOIprefix}{doi:}
\providecommand{\ArXivprefix}{arXiv:}
\providecommand{\URLprefix}{URL: }
\providecommand{\Pubmedprefix}{pmid:}
\providecommand{\doi}[1]{\href{http://dx.doi.org/#1}{\path{#1}}}
\providecommand{\Pubmed}[1]{\href{pmid:#1}{\path{#1}}}
\providecommand{\BIBand}{and}
\providecommand{\bibinfo}[2]{#2}
\ifx\xfnm\undefined \def\xfnm[#1]{\unskip,\space#1}\fi
\makeatletter\def\@biblabel#1{#1.}\makeatother
\bibitem[{Moore and Teaney(2005)}]{Moore:2004tg}
\bibinfo{author}{Moore\xfnm[ G.D.]}, \bibinfo{author}{Teaney\xfnm[ D.]}.
\newblock \bibinfo{title}{{How much do heavy quarks thermalize in a heavy ion
  collision?}}
\newblock \emph{\bibinfo{journal}{Phys Rev C}};
  \bibinfo{year}{2005};\bibinfo{volume}{71}:\bibinfo{pages}{064904}.
\newblock \DOIprefix\doi{10.1103/PhysRevC.71.064904};
  \href{http://arxiv.org/abs/hep-ph/0412346}{\tt arXiv:hep-ph/0412346}.
\bibitem[{Gubser(2006)}]{Gubser:2006bz}
\bibinfo{author}{Gubser\xfnm[ S.S.]}.
\newblock \bibinfo{title}{{Drag force in AdS/CFT}}.
\newblock \emph{\bibinfo{journal}{Phys Rev D}};
  \bibinfo{year}{2006};\bibinfo{volume}{74}:\bibinfo{pages}{126005}.
\newblock \DOIprefix\doi{10.1103/PhysRevD.74.126005};
  \href{http://arxiv.org/abs/hep-th/0605182}{\tt arXiv:hep-th/0605182}.
\bibitem[{Das et~al.(2010)Das, Alam and Mohanty}]{Das:2010tj}
\bibinfo{author}{Das\xfnm[ S.K.]}, \bibinfo{author}{Alam\xfnm[ J.e.]},
  \bibinfo{author}{Mohanty\xfnm[ P.]}.
\newblock \bibinfo{title}{{Dragging Heavy Quarks in Quark Gluon Plasma at the
  Large Hadron Collider}}.
\newblock \emph{\bibinfo{journal}{Phys Rev C}};
  \bibinfo{year}{2010};\bibinfo{volume}{82}:\bibinfo{pages}{014908}.
\newblock \DOIprefix\doi{10.1103/PhysRevC.82.014908};
  \href{http://arxiv.org/abs/1003.5508}{\tt arXiv:1003.5508}.
\bibitem[{Akamatsu et~al.(2009)Akamatsu, Hatsuda and Hirano}]{Akamatsu:2008ge}
\bibinfo{author}{Akamatsu\xfnm[ Y.]}, \bibinfo{author}{Hatsuda\xfnm[ T.]},
  \bibinfo{author}{Hirano\xfnm[ T.]}.
\newblock \bibinfo{title}{{Heavy Quark Diffusion with Relativistic Langevin
  Dynamics in the Quark-Gluon Fluid}}.
\newblock \emph{\bibinfo{journal}{Phys Rev C}};
  \bibinfo{year}{2009};\bibinfo{volume}{79}:\bibinfo{pages}{054907}.
\newblock \DOIprefix\doi{10.1103/PhysRevC.79.054907};
  \href{http://arxiv.org/abs/0809.1499}{\tt arXiv:0809.1499}.
\bibitem[{Banerjee et~al.(2012)Banerjee, Datta, Gavai and
  Majumdar}]{Banerjee:2011ra}
\bibinfo{author}{Banerjee\xfnm[ D.]}, \bibinfo{author}{Datta\xfnm[ S.]},
  \bibinfo{author}{Gavai\xfnm[ R.]}, \bibinfo{author}{Majumdar\xfnm[ P.]}.
\newblock \bibinfo{title}{{Heavy Quark Momentum Diffusion Coefficient from
  Lattice QCD}}.
\newblock \emph{\bibinfo{journal}{Phys Rev D}};
  \bibinfo{year}{2012};\bibinfo{volume}{85}:\bibinfo{pages}{014510}.
\newblock \DOIprefix\doi{10.1103/PhysRevD.85.014510};
  \href{http://arxiv.org/abs/1109.5738}{\tt arXiv:1109.5738}.
\bibitem[{Ding et~al.(2014)Ding, Francis, Kaczmarek, Karsch, Satz and
  S\"oldner}]{Ding:2012iy}
\bibinfo{author}{Ding\xfnm[ H.T.]}, \bibinfo{author}{Francis\xfnm[ A.]},
  \bibinfo{author}{Kaczmarek\xfnm[ O.]}, \bibinfo{author}{Karsch\xfnm[ F.]},
  \bibinfo{author}{Satz\xfnm[ H.]}, \bibinfo{author}{S\"oldner\xfnm[ W.]}.
\newblock \bibinfo{title}{{Charmonium dissociation and heavy quark transport in
  hot quenched lattice QCD}}.
\newblock \emph{\bibinfo{journal}{EPJ Web Conf}};
  \bibinfo{year}{2014};\bibinfo{volume}{70}:\bibinfo{pages}{00061}.
\newblock \DOIprefix\doi{10.1051/epjconf/20147000061};
  \href{http://arxiv.org/abs/1210.0292}{\tt arXiv:1210.0292}.
\bibitem[{van Hees et~al.(2008{\natexlab{a}})van Hees, Mannarelli, Greco and
  Rapp}]{vanHees:2007me}
\bibinfo{author}{van Hees\xfnm[ H.]}, \bibinfo{author}{Mannarelli\xfnm[ M.]},
  \bibinfo{author}{Greco\xfnm[ V.]}, \bibinfo{author}{Rapp\xfnm[ R.]}.
\newblock \bibinfo{title}{{Nonperturbative heavy-quark diffusion in the
  quark-gluon plasma}}.
\newblock \emph{\bibinfo{journal}{Phys Rev Lett}};
  \bibinfo{year}{2008}{\natexlab{a}};\bibinfo{volume}{100}:\bibinfo{pages}{192301}.
\newblock \DOIprefix\doi{10.1103/PhysRevLett.100.192301};
  \href{http://arxiv.org/abs/0709.2884}{\tt arXiv:0709.2884}.
\bibitem[{Svetitsky(1988)}]{PhysRevD.37.2484}
\bibinfo{author}{Svetitsky\xfnm[ B.]}.
\newblock \bibinfo{title}{Diffusion of charmed quarks in the quark-gluon
  plasma}.
\newblock \emph{\bibinfo{journal}{Phys Rev D}};
  \bibinfo{year}{1988};\bibinfo{volume}{37}:\bibinfo{pages}{2484--2491}.
\newblock \URLprefix \url{https://link.aps.org/doi/10.1103/PhysRevD.37.2484};
  \DOIprefix\doi{10.1103/PhysRevD.37.2484}.
\bibitem[{Dong and Greco(2019)}]{dong2019heavy}
\bibinfo{author}{Dong\xfnm[ X.]}, \bibinfo{author}{Greco\xfnm[ V.]}.
\newblock \bibinfo{title}{Heavy quark production and properties of quark--gluon
  plasma}.
\newblock \emph{\bibinfo{journal}{Progress in Particle and Nuclear Physics}};
  \bibinfo{year}{2019};\bibinfo{volume}{104}:\bibinfo{pages}{97--141}.
\bibitem[{van Hees et~al.(2006)van Hees, Greco and Rapp}]{PhysRevC.73.034913}
\bibinfo{author}{van Hees\xfnm[ H.]}, \bibinfo{author}{Greco\xfnm[ V.]},
  \bibinfo{author}{Rapp\xfnm[ R.]}.
\newblock \bibinfo{title}{Heavy-quark probes of the quark-gluon plasma and
  interpretation of recent data taken at the bnl relativistic heavy ion
  collider}.
\newblock \emph{\bibinfo{journal}{Phys Rev C}};
  \bibinfo{year}{2006};\bibinfo{volume}{73}:\bibinfo{pages}{034913}.
\newblock \URLprefix \url{https://link.aps.org/doi/10.1103/PhysRevC.73.034913};
  \DOIprefix\doi{10.1103/PhysRevC.73.034913}.
\bibitem[{Das et~al.(2015)Das, Scardina, Plumari and Greco}]{das2015toward}
\bibinfo{author}{Das\xfnm[ S.K.]}, \bibinfo{author}{Scardina\xfnm[ F.]},
  \bibinfo{author}{Plumari\xfnm[ S.]}, \bibinfo{author}{Greco\xfnm[ V.]}.
\newblock \bibinfo{title}{Toward a solution to the raa and v2 puzzle for heavy
  quarks}.
\newblock \emph{\bibinfo{journal}{Physics Letters B}};
  \bibinfo{year}{2015};\bibinfo{volume}{747}:\bibinfo{pages}{260--264}.
\bibitem[{van Hees et~al.(2008{\natexlab{b}})van Hees, Mannarelli, Greco and
  Rapp}]{PhysRevLett.100.192301}
\bibinfo{author}{van Hees\xfnm[ H.]}, \bibinfo{author}{Mannarelli\xfnm[ M.]},
  \bibinfo{author}{Greco\xfnm[ V.]}, \bibinfo{author}{Rapp\xfnm[ R.]}.
\newblock \bibinfo{title}{Nonperturbative heavy-quark diffusion in the
  quark-gluon plasma}.
\newblock \emph{\bibinfo{journal}{Phys Rev Lett}};
  \bibinfo{year}{2008}{\natexlab{b}};\bibinfo{volume}{100}:\bibinfo{pages}{192301}.
\newblock \URLprefix
  \url{https://link.aps.org/doi/10.1103/PhysRevLett.100.192301};
  \DOIprefix\doi{10.1103/PhysRevLett.100.192301}.
\bibitem[{Acharya et~al.(2022{\natexlab{a}})}]{ALICE:2021rxa}
\bibinfo{author}{Acharya\xfnm[ S.]}, et~al. (\bibinfo{collaboration}{ALICE}).
\newblock \bibinfo{title}{{Prompt D$^{0}$, D$^{+}$, and D$^{*+}$ production in
  Pb\textendash{}Pb collisions at $ \sqrt{s_{\mathrm{NN}}} $ = 5.02 TeV}}.
\newblock \emph{\bibinfo{journal}{JHEP}};
  \bibinfo{year}{2022}{\natexlab{a}};\bibinfo{volume}{01}:\bibinfo{pages}{174}.
\newblock \DOIprefix\doi{10.1007/JHEP01(2022)174};
  \href{http://arxiv.org/abs/2110.09420}{\tt arXiv:2110.09420}.
\bibitem[{Adamczyk et~al.(2017{\natexlab{a}})}]{STAR:2017kkh}
\bibinfo{author}{Adamczyk\xfnm[ L.]}, et~al. (\bibinfo{collaboration}{STAR}).
\newblock \bibinfo{title}{{Measurement of $D^0$ Azimuthal Anisotropy at
  Midrapidity in Au+Au Collisions at $\sqrt{s_{NN}}$=200 GeV}}.
\newblock \emph{\bibinfo{journal}{Phys Rev Lett}};
  \bibinfo{year}{2017}{\natexlab{a}};\bibinfo{volume}{118}(\bibinfo{number}{21}):\bibinfo{pages}{212301}.
\newblock \DOIprefix\doi{10.1103/PhysRevLett.118.212301};
  \href{http://arxiv.org/abs/1701.06060}{\tt arXiv:1701.06060}.
\bibitem[{Liang and Wang(2005{\natexlab{a}})}]{Liang:2004ph}
\bibinfo{author}{Liang\xfnm[ Z.T.]}, \bibinfo{author}{Wang\xfnm[ X.N.]}.
\newblock \bibinfo{title}{{Globally polarized quark-gluon plasma in non-central
  A+A collisions}}.
\newblock \emph{\bibinfo{journal}{Phys Rev Lett}};
  \bibinfo{year}{2005}{\natexlab{a}};\bibinfo{volume}{94}:\bibinfo{pages}{102301}.
\newblock \DOIprefix\doi{10.1103/PhysRevLett.94.102301};
  \href{http://arxiv.org/abs/nucl-th/0410079}{\tt arXiv:nucl-th/0410079};
  \bibinfo{note}{[Erratum: Phys.Rev.Lett. 96, 039901 (2006)]}.
\bibitem[{Liang and Wang(2005{\natexlab{b}})}]{Liang:2004xn}
\bibinfo{author}{Liang\xfnm[ Z.T.]}, \bibinfo{author}{Wang\xfnm[ X.N.]}.
\newblock \bibinfo{title}{{Spin alignment of vector mesons in non-central A+A
  collisions}}.
\newblock \emph{\bibinfo{journal}{Phys Lett B}};
  \bibinfo{year}{2005}{\natexlab{b}};\bibinfo{volume}{629}:\bibinfo{pages}{20--26}.
\newblock \DOIprefix\doi{10.1016/j.physletb.2005.09.060};
  \href{http://arxiv.org/abs/nucl-th/0411101}{\tt arXiv:nucl-th/0411101}.
\bibitem[{Adamczyk et~al.(2017{\natexlab{b}})}]{STAR:2017ckg}
\bibinfo{author}{Adamczyk\xfnm[ L.]}, et~al. (\bibinfo{collaboration}{STAR}).
\newblock \bibinfo{title}{{Global $\Lambda$ hyperon polarization in nuclear
  collisions: evidence for the most vortical fluid}}.
\newblock \emph{\bibinfo{journal}{Nature}};
  \bibinfo{year}{2017}{\natexlab{b}};\bibinfo{volume}{548}:\bibinfo{pages}{62--65}.
\newblock \DOIprefix\doi{10.1038/nature23004};
  \href{http://arxiv.org/abs/1701.06657}{\tt arXiv:1701.06657}.
\bibitem[{Adam et~al.(2018{\natexlab{a}})}]{STAR:2018pps}
\bibinfo{author}{Adam\xfnm[ J.]}, et~al. (\bibinfo{collaboration}{STAR}).
\newblock \bibinfo{title}{{Improved measurement of the longitudinal spin
  transfer to $\Lambda$ and $\bar \Lambda$ hyperons in polarized proton-proton
  collisions at $\sqrt s$ = 200 GeV}}.
\newblock \emph{\bibinfo{journal}{Phys Rev D}};
  \bibinfo{year}{2018}{\natexlab{a}};\bibinfo{volume}{98}(\bibinfo{number}{11}):\bibinfo{pages}{112009}.
\newblock \DOIprefix\doi{10.1103/PhysRevD.98.112009};
  \href{http://arxiv.org/abs/1808.07634}{\tt arXiv:1808.07634}.
\bibitem[{Adam et~al.(2018{\natexlab{b}})}]{STAR:2018fqv}
\bibinfo{author}{Adam\xfnm[ J.]}, et~al. (\bibinfo{collaboration}{STAR}).
\newblock \bibinfo{title}{{Transverse spin transfer to $\Lambda$ and
  $\bar{\Lambda}$ hyperons in polarized proton-proton collisions at
  $\sqrt{s}=200\,\mathrm{GeV}$}}.
\newblock \emph{\bibinfo{journal}{Phys Rev D}};
  \bibinfo{year}{2018}{\natexlab{b}};\bibinfo{volume}{98}(\bibinfo{number}{9}):\bibinfo{pages}{091103}.
\newblock \DOIprefix\doi{10.1103/PhysRevD.98.091103};
  \href{http://arxiv.org/abs/1808.08000}{\tt arXiv:1808.08000}.
\bibitem[{Niida(2019)}]{Niida:2018hfw}
\bibinfo{author}{Niida\xfnm[ T.]} (\bibinfo{collaboration}{STAR}).
\newblock \bibinfo{title}{{Global and local polarization of $\Lambda$ hyperons
  in Au+Au collisions at 200 GeV from STAR}}.
\newblock \emph{\bibinfo{journal}{Nucl Phys}};
  \bibinfo{year}{2019};\bibinfo{volume}{A982}:\bibinfo{pages}{511--514}.
\newblock \DOIprefix\doi{10.1016/j.nuclphysa.2018.08.034};
  \href{http://arxiv.org/abs/1808.10482}{\tt arXiv:1808.10482}.
\bibitem[{Adam et~al.(2018{\natexlab{c}})}]{STAR:2018gyt}
\bibinfo{author}{Adam\xfnm[ J.]}, et~al. (\bibinfo{collaboration}{STAR}).
\newblock \bibinfo{title}{{Global polarization of $\Lambda$ hyperons in Au+Au
  collisions at $\sqrt{s_{_{NN}}}$ = 200 GeV}}.
\newblock \emph{\bibinfo{journal}{Phys Rev C}};
  \bibinfo{year}{2018}{\natexlab{c}};\bibinfo{volume}{98}:\bibinfo{pages}{014910}.
\newblock \DOIprefix\doi{10.1103/PhysRevC.98.014910};
  \href{http://arxiv.org/abs/1805.04400}{\tt arXiv:1805.04400}.
\bibitem[{Acharya et~al.(2020{\natexlab{a}})}]{ALICE:2019aid}
\bibinfo{author}{Acharya\xfnm[ S.]}, et~al. (\bibinfo{collaboration}{ALICE}).
\newblock \bibinfo{title}{{Evidence of Spin-Orbital Angular Momentum
  Interactions in Relativistic Heavy-Ion Collisions}}.
\newblock \emph{\bibinfo{journal}{Phys Rev Lett}};
  \bibinfo{year}{2020}{\natexlab{a}};\bibinfo{volume}{125}(\bibinfo{number}{1}):\bibinfo{pages}{012301}.
\newblock \DOIprefix\doi{10.1103/PhysRevLett.125.012301};
  \href{http://arxiv.org/abs/1910.14408}{\tt arXiv:1910.14408}.
\bibitem[{Adam et~al.(2019)}]{STAR:2019erd}
\bibinfo{author}{Adam\xfnm[ J.]}, et~al. (\bibinfo{collaboration}{STAR}).
\newblock \bibinfo{title}{{Polarization of $\Lambda$ ($\bar{\Lambda}$) hyperons
  along the beam direction in Au+Au collisions at $\sqrt{s_{_{NN}}}$ = 200
  GeV}}.
\newblock \emph{\bibinfo{journal}{Phys Rev Lett}};
  \bibinfo{year}{2019};\bibinfo{volume}{123}(\bibinfo{number}{13}):\bibinfo{pages}{132301}.
\newblock \DOIprefix\doi{10.1103/PhysRevLett.123.132301};
  \href{http://arxiv.org/abs/1905.11917}{\tt arXiv:1905.11917}.
\bibitem[{Acharya et~al.(2020{\natexlab{b}})}]{ALICE:2019onw}
\bibinfo{author}{Acharya\xfnm[ S.]}, et~al. (\bibinfo{collaboration}{ALICE}).
\newblock \bibinfo{title}{{Global polarization of $\Lambda \bar \Lambda$
  hyperons in Pb-Pb collisions at $\sqrt {s_{NN}}$ = 2.76 and 5.02 TeV}}.
\newblock \emph{\bibinfo{journal}{Phys Rev C}};
  \bibinfo{year}{2020}{\natexlab{b}};\bibinfo{volume}{101}(\bibinfo{number}{4}):\bibinfo{pages}{044611}.
\newblock \DOIprefix\doi{10.1103/PhysRevC.101.044611};
  \href{http://arxiv.org/abs/1909.01281}{\tt arXiv:1909.01281};
  \bibinfo{note}{[Erratum: Phys.Rev.C 105, 029902 (2022)]}.
\bibitem[{Singha(2021)}]{Singha:2020qns}
\bibinfo{author}{Singha\xfnm[ S.]} (\bibinfo{collaboration}{STAR}).
\newblock \bibinfo{title}{{Measurement of global spin alignment of $K^{*0}$and
  $\phi$ vector mesons using the STAR detector at RHIC}}.
\newblock \emph{\bibinfo{journal}{Nucl Phys A}};
  \bibinfo{year}{2021};\bibinfo{volume}{1005}:\bibinfo{pages}{121733}.
\newblock \DOIprefix\doi{10.1016/j.nuclphysa.2020.121733};
  \href{http://arxiv.org/abs/2002.07427}{\tt arXiv:2002.07427}.
\bibitem[{Chen et~al.(2020)Chen, Liang, Song and Wei}]{Chen:2020pty}
\bibinfo{author}{Chen\xfnm[ K.b.]}, \bibinfo{author}{Liang\xfnm[ Z.t.]},
  \bibinfo{author}{Song\xfnm[ Y.k.]}, \bibinfo{author}{Wei\xfnm[ S.y.]}.
\newblock \bibinfo{title}{{Spin alignment of vector mesons in high energy $pp$
  collisions}}.
\newblock \emph{\bibinfo{journal}{Phys Rev D}};
  \bibinfo{year}{2020};\bibinfo{volume}{102}(\bibinfo{number}{3}):\bibinfo{pages}{034001}.
\newblock \DOIprefix\doi{10.1103/PhysRevD.102.034001};
  \href{http://arxiv.org/abs/2002.09890}{\tt arXiv:2002.09890}.
\bibitem[{Adam et~al.(2021)}]{STAR:2020xbm}
\bibinfo{author}{Adam\xfnm[ J.]}, et~al. (\bibinfo{collaboration}{STAR}).
\newblock \bibinfo{title}{{Global Polarization of $\Xi$ and $\Omega$ Hyperons
  in Au+Au Collisions at $\sqrt {s_{NN}}$ = 200 GeV}}.
\newblock \emph{\bibinfo{journal}{Phys Rev Lett}};
  \bibinfo{year}{2021};\bibinfo{volume}{126}(\bibinfo{number}{16}):\bibinfo{pages}{162301}.
\newblock \DOIprefix\doi{10.1103/PhysRevLett.126.162301};
  \href{http://arxiv.org/abs/2012.13601}{\tt arXiv:2012.13601}.
\bibitem[{Acharya et~al.(2022{\natexlab{b}})}]{ALICE:2021pzu}
\bibinfo{author}{Acharya\xfnm[ S.]}, et~al. (\bibinfo{collaboration}{ALICE}).
\newblock \bibinfo{title}{{Polarization of $\Lambda$ and $\bar \Lambda$
  Hyperons along the Beam Direction in Pb-Pb Collisions at $\sqrt
  {s_{NN}}$=5.02\,\,TeV}}.
\newblock \emph{\bibinfo{journal}{Phys Rev Lett}};
  \bibinfo{year}{2022}{\natexlab{b}};\bibinfo{volume}{128}(\bibinfo{number}{17}):\bibinfo{pages}{172005}.
\newblock \DOIprefix\doi{10.1103/PhysRevLett.128.172005};
  \href{http://arxiv.org/abs/2107.11183}{\tt arXiv:2107.11183}.
\bibitem[{Abdallah et~al.(2021)}]{STAR:2021beb}
\bibinfo{author}{Abdallah\xfnm[ M.S.]}, et~al. (\bibinfo{collaboration}{STAR}).
\newblock \bibinfo{title}{{Global $\Lambda$-hyperon polarization in Au+Au
  collisions at $\sqrt {s_{NN}}$=3~GeV}}.
\newblock \emph{\bibinfo{journal}{Phys Rev C}};
  \bibinfo{year}{2021};\bibinfo{volume}{104}(\bibinfo{number}{6}):\bibinfo{pages}{L061901}.
\newblock \DOIprefix\doi{10.1103/PhysRevC.104.L061901};
  \href{http://arxiv.org/abs/2108.00044}{\tt arXiv:2108.00044}.
\bibitem[{Mohanty et~al.(2021)Mohanty, Kundu, Singha and
  Singh}]{Mohanty:2021vbt}
\bibinfo{author}{Mohanty\xfnm[ B.]}, \bibinfo{author}{Kundu\xfnm[ S.]},
  \bibinfo{author}{Singha\xfnm[ S.]}, \bibinfo{author}{Singh\xfnm[ R.]}.
\newblock \bibinfo{title}{{Spin alignment measurement of vector mesons produced
  in high energy collisions}}.
\newblock \emph{\bibinfo{journal}{Mod Phys Lett A}};
  \bibinfo{year}{2021};\bibinfo{volume}{36}(\bibinfo{number}{39}):\bibinfo{pages}{2130026}.
\newblock \DOIprefix\doi{10.1142/S0217732321300263};
  \href{http://arxiv.org/abs/2112.04816}{\tt arXiv:2112.04816}.
\bibitem[{Abdallah et~al.(2023)}]{STAR:2022fan}
\bibinfo{author}{Abdallah\xfnm[ M.S.]}, et~al. (\bibinfo{collaboration}{STAR}).
\newblock \bibinfo{title}{{Pattern of global spin alignment of
  \ensuremath{\phi} and K$^{*0}$ mesons in heavy-ion collisions}}.
\newblock \emph{\bibinfo{journal}{Nature}};
  \bibinfo{year}{2023};\bibinfo{volume}{614}(\bibinfo{number}{7947}):\bibinfo{pages}{244--248}.
\newblock \DOIprefix\doi{10.1038/s41586-022-05557-5};
  \href{http://arxiv.org/abs/2204.02302}{\tt arXiv:2204.02302}.
\bibitem[{Abdulhamid et~al.(2023)}]{STAR:2023eck}
\bibinfo{author}{Abdulhamid\xfnm[ M.]}, et~al. (\bibinfo{collaboration}{STAR}).
\newblock \bibinfo{title}{{Hyperon Polarization along the Beam Direction
  Relative to the Second and Third Harmonic Event Planes in Isobar Collisions
  at sNN=200\,\,GeV}}.
\newblock \emph{\bibinfo{journal}{Phys Rev Lett}};
  \bibinfo{year}{2023};\bibinfo{volume}{131}(\bibinfo{number}{20}):\bibinfo{pages}{202301}.
\newblock \DOIprefix\doi{10.1103/PhysRevLett.131.202301};
  \href{http://arxiv.org/abs/2303.09074}{\tt arXiv:2303.09074}.
\bibitem[{Debye(1929)}]{Debye1929}
\bibinfo{author}{Debye\xfnm[ P.]}.
\newblock \bibinfo{title}{Polar Molecules}.
\newblock \bibinfo{publisher}{Chemical Catalog Co New York};
  \bibinfo{year}{1929}.
\bibitem[{Perrin(1934)}]{Perrin1934}
\bibinfo{author}{Perrin\xfnm[ F.]}.
\newblock \bibinfo{title}{Mouvement brownien d'un ellipsoide - i. dispersion
  diélectrique pour des molécules ellipsoidales}.
\newblock \emph{\bibinfo{journal}{J Phys Radium}};
  \bibinfo{year}{1934};\bibinfo{volume}{5}(\bibinfo{number}{10}):\bibinfo{pages}{497--511}.
\newblock \URLprefix \url{https://doi.org/10.1051/jphysrad:01934005010049700};
  \DOIprefix\doi{10.1051/jphysrad:01934005010049700}.
\bibitem[{Landau and Lifstatz(1935)}]{Landau:1935qbc}
\bibinfo{author}{Landau\xfnm[ L.D.]}, \bibinfo{author}{Lifstatz\xfnm[ E.]}.
\newblock \bibinfo{title}{{On the Theory of the Dispersion of Magnetic
  Permeability in Ferromagnetic Bodies}}.
\newblock \emph{\bibinfo{journal}{Phys Z Sowjetunion}};
  \bibinfo{year}{1935};\bibinfo{volume}{8}.
\newblock \DOIprefix\doi{10.1016/b978-0-08-010586-4.50023-7}.
\bibitem[{Furry(1957)}]{Furry1957}
\bibinfo{author}{Furry\xfnm[ W.H.]}.
\newblock \bibinfo{title}{Isotropic rotational brownian motion}.
\newblock \emph{\bibinfo{journal}{Phys Rev}};
  \bibinfo{year}{1957};\bibinfo{volume}{107}:\bibinfo{pages}{7--13}.
\newblock \URLprefix \url{https://link.aps.org/doi/10.1103/PhysRev.107.7};
  \DOIprefix\doi{10.1103/PhysRev.107.7}.
\bibitem[{Favro(1960)}]{Favro1960}
\bibinfo{author}{Favro\xfnm[ L.D.]}.
\newblock \bibinfo{title}{Theory of the rotational brownian motion of a free
  rigid body}.
\newblock \emph{\bibinfo{journal}{Phys Rev}};
  \bibinfo{year}{1960};\bibinfo{volume}{119}:\bibinfo{pages}{53--62}.
\newblock \URLprefix \url{https://link.aps.org/doi/10.1103/PhysRev.119.53};
  \DOIprefix\doi{10.1103/PhysRev.119.53}.
\bibitem[{Evanov(1964)}]{Evanov1964}
\bibinfo{author}{Evanov\xfnm[ E.N.]}.
\newblock \bibinfo{title}{Theory of rotational brownian motion}.
\newblock \emph{\bibinfo{journal}{JETP}};
  \bibinfo{year}{1964};\bibinfo{volume}{18}(\bibinfo{number}{4}):\bibinfo{pages}{1041}.
\bibitem[{Hubbard(1972)}]{Hubbard1972}
\bibinfo{author}{Hubbard\xfnm[ P.S.]}.
\newblock \bibinfo{title}{Rotational brownian motion}.
\newblock \emph{\bibinfo{journal}{Phys Rev A}};
  \bibinfo{year}{1972};\bibinfo{volume}{6}:\bibinfo{pages}{2421--2433}.
\newblock \URLprefix \url{https://link.aps.org/doi/10.1103/PhysRevA.6.2421};
  \DOIprefix\doi{10.1103/PhysRevA.6.2421}.
\bibitem[{Valiev and Ivanov(1973)}]{Valiev_1973}
\bibinfo{author}{Valiev\xfnm[ K.A.]}, \bibinfo{author}{Ivanov\xfnm[ E.N.]}.
\newblock \bibinfo{title}{Rotational brownian motion}.
\newblock \emph{\bibinfo{journal}{Soviet Physics Uspekhi}};
  \bibinfo{year}{1973};\bibinfo{volume}{16}(\bibinfo{number}{1}):\bibinfo{pages}{1}.
\newblock \URLprefix \url{https://dx.doi.org/10.1070/PU1973v016n01ABEH005145};
  \DOIprefix\doi{10.1070/PU1973v016n01ABEH005145}.
\bibitem[{Coffey(1980)}]{COFFEY1980}
\bibinfo{author}{Coffey\xfnm[ W.]}.
\newblock \bibinfo{title}{Rotational and translational brownian motion}.
\newblock \emph{\bibinfo{journal}{Advances in Molecular Relaxation and
  Interaction Processes}};
  \bibinfo{year}{1980};\bibinfo{volume}{17}(\bibinfo{number}{3}):\bibinfo{pages}{169--337}.
\newblock \URLprefix
  \url{https://www.sciencedirect.com/science/article/pii/0378448780800446};
  \DOIprefix\doi{https://doi.org/10.1016/0378-4487(80)80044-6}.
\bibitem[{Coffey et~al.(2003)Coffey, Kalmykov and Titov}]{Coffey_2003}
\bibinfo{author}{Coffey\xfnm[ W.T.]}, \bibinfo{author}{Kalmykov\xfnm[ Y.P.]},
  \bibinfo{author}{Titov\xfnm[ S.V.]}.
\newblock \bibinfo{title}{Langevin equation method for the rotational brownian
  motion and orientational relaxation in liquids: Ii. symmetrical top
  molecules}.
\newblock \emph{\bibinfo{journal}{Journal of Physics A: Mathematical and
  General}};
  \bibinfo{year}{2003};\bibinfo{volume}{36}(\bibinfo{number}{18}):\bibinfo{pages}{4947}.
\newblock \URLprefix \url{https://dx.doi.org/10.1088/0305-4470/36/18/301};
  \DOIprefix\doi{10.1088/0305-4470/36/18/301}.
\bibitem[{Kubo(1970)}]{KuboSPINFIRST}
\bibinfo{author}{Kubo Ryogo A1~Hashitsume\xfnm[ N.]}.
\newblock \bibinfo{title}{Brownian motion of spins}.
\newblock \emph{\bibinfo{journal}{Progress of Theoretical Physics Supplement}};
  \bibinfo{year}{1970};\bibinfo{volume}{46}:\bibinfo{pages}{210–220}.
\newblock \URLprefix \url{https://doi.org/10.1143/PTPS.46.210};
  \DOIprefix\doi{10.1143/PTPS.46.210}.
\bibitem[{Liu et~al.(2024)Liu, Bai, Zheng, Huang and Chen}]{Liu:2024hii}
\bibinfo{author}{Liu\xfnm[ Z.]}, \bibinfo{author}{Bai\xfnm[ Y.]},
  \bibinfo{author}{Zheng\xfnm[ S.]}, \bibinfo{author}{Huang\xfnm[ A.]},
  \bibinfo{author}{Chen\xfnm[ B.]}.
\newblock \bibinfo{title}{{Exploring spin polarization of heavy quarks in
  magnetic fields and hot medium}}.
\newblock \emph{\bibinfo{journal}{Phys Rev C}};
  \bibinfo{year}{2024};\bibinfo{volume}{110}(\bibinfo{number}{3}):\bibinfo{pages}{034910}.
\newblock \DOIprefix\doi{10.1103/PhysRevC.110.034910};
  \href{http://arxiv.org/abs/2404.02032}{\tt arXiv:2404.02032}.
\bibitem[{Acharya et~al.(2025)}]{ALICE:2025cdf}
\bibinfo{author}{Acharya\xfnm[ S.]}, et~al. (\bibinfo{collaboration}{ALICE}).
\newblock \bibinfo{title}{{First measurement of D$^{*+}$ vector meson spin
  alignment in Pb{\textendash}Pb collisions at
  $\sqrt{{s}_{\text{NN}}}={5}.0{2}$ TeV}}.
\newblock \emph{\bibinfo{journal}{JHEP}};
  \bibinfo{year}{2025};\bibinfo{volume}{10}:\bibinfo{pages}{094}.
\newblock \DOIprefix\doi{10.1007/JHEP10(2025)094};
  \href{http://arxiv.org/abs/2504.00714}{\tt arXiv:2504.00714}.
\bibitem[{Gilbert(2004)}]{Gilbert_2004}
\bibinfo{author}{Gilbert\xfnm[ T.]}.
\newblock \bibinfo{title}{A phenomenological theory of damping in ferromagnetic
  materials}.
\newblock \emph{\bibinfo{journal}{IEEE Transactions on Magnetics}};
  \bibinfo{year}{2004};\bibinfo{volume}{40}(\bibinfo{number}{6}):\bibinfo{pages}{3443--3449}.
\newblock \DOIprefix\doi{10.1109/TMAG.2004.836740}.
\bibitem[{Nishino and Miyashita(2015{\natexlab{a}})}]{Nishino_2015}
\bibinfo{author}{Nishino\xfnm[ M.]}, \bibinfo{author}{Miyashita\xfnm[ S.]}.
\newblock \bibinfo{title}{Realization of the thermal equilibrium in
  inhomogeneous magnetic systems by the landau-lifshitz-gilbert equation with
  stochastic noise, and its dynamical aspects}.
\newblock \emph{\bibinfo{journal}{Phys Rev B}};
  \bibinfo{year}{2015}{\natexlab{a}};\bibinfo{volume}{91}:\bibinfo{pages}{134411}.
\newblock \URLprefix \url{https://link.aps.org/doi/10.1103/PhysRevB.91.134411};
  \DOIprefix\doi{10.1103/PhysRevB.91.134411}.
\bibitem[{Meo et~al.(2022)Meo, Cronshaw, Jenkins, Lees and Evans}]{Meo_2023}
\bibinfo{author}{Meo\xfnm[ A.]}, \bibinfo{author}{Cronshaw\xfnm[ C.E.]},
  \bibinfo{author}{Jenkins\xfnm[ S.]}, \bibinfo{author}{Lees\xfnm[ A.]},
  \bibinfo{author}{Evans\xfnm[ R.F.L.]}.
\newblock \bibinfo{title}{Spin-transfer and spin-orbit torques in the
  landau–lifshitz–gilbert equation}.
\newblock \emph{\bibinfo{journal}{Journal of Physics: Condensed Matter}};
  \bibinfo{year}{2022};\bibinfo{volume}{35}(\bibinfo{number}{2}):\bibinfo{pages}{025801}.
\newblock \URLprefix \url{https://dx.doi.org/10.1088/1361-648X/ac9c80};
  \DOIprefix\doi{10.1088/1361-648X/ac9c80}.
\bibitem[{Leader(2001)}]{Leader_2001}
\bibinfo{author}{Leader\xfnm[ E.]}.
\newblock \bibinfo{title}{Spin in Particle Physics}.
\newblock Cambridge Monographs on Particle Physics, Nuclear Physics and
  Cosmology; \bibinfo{publisher}{Cambridge University Press};
  \bibinfo{year}{2001}.
\bibitem[{Risken and Risken(1996)}]{risken1996fokker}
\bibinfo{author}{Risken\xfnm[ H.]}, \bibinfo{author}{Risken\xfnm[ H.]}.
\newblock \bibinfo{title}{Fokker-planck equation}.
\newblock \bibinfo{publisher}{Springer}; \bibinfo{year}{1996}.
\bibitem[{Balakrishnan(0083)}]{Balakrishnan}
\bibinfo{author}{Balakrishnan\xfnm[ V.]}.
\newblock \bibinfo{title}{Elements of nonequilibrium statistical mechanics};
  vol.~\bibinfo{volume}{3}.
\newblock \bibinfo{publisher}{Ane Books}; \bibinfo{year}{2008/3}.
\bibitem[{Nishino and Miyashita(2015{\natexlab{b}})}]{PhysRevB.91.134411}
\bibinfo{author}{Nishino\xfnm[ M.]}, \bibinfo{author}{Miyashita\xfnm[ S.]}.
\newblock \bibinfo{title}{Realization of the thermal equilibrium in
  inhomogeneous magnetic systems by the landau-lifshitz-gilbert equation with
  stochastic noise, and its dynamical aspects}.
\newblock \emph{\bibinfo{journal}{Phys Rev B}};
  \bibinfo{year}{2015}{\natexlab{b}};\bibinfo{volume}{91}:\bibinfo{pages}{134411}.
\newblock \URLprefix \url{https://link.aps.org/doi/10.1103/PhysRevB.91.134411};
  \DOIprefix\doi{10.1103/PhysRevB.91.134411}.
\bibitem[{Taniguchi and Imamura(2011)}]{PhysRevB.83.054432}
\bibinfo{author}{Taniguchi\xfnm[ T.]}, \bibinfo{author}{Imamura\xfnm[ H.]}.
\newblock \bibinfo{title}{Thermally assisted spin transfer torque switching in
  synthetic free layers}.
\newblock \emph{\bibinfo{journal}{Phys Rev B}};
  \bibinfo{year}{2011};\bibinfo{volume}{83}:\bibinfo{pages}{054432}.
\newblock \URLprefix \url{https://link.aps.org/doi/10.1103/PhysRevB.83.054432};
  \DOIprefix\doi{10.1103/PhysRevB.83.054432}.
\bibitem[{Livi and Politi(2017)}]{Livi_Politi_2017}
\bibinfo{author}{Livi\xfnm[ R.]}, \bibinfo{author}{Politi\xfnm[ P.]}.
\newblock \bibinfo{title}{Nonequilibrium Statistical Physics: A Modern
  Perspective}.
\newblock \bibinfo{publisher}{Cambridge University Press};
  \bibinfo{year}{2017}.
\bibitem[{Garc\'{\i}a-Palacios and L\'azaro(1998)}]{PhysRevB.58.14937}
\bibinfo{author}{Garc\'{\i}a-Palacios\xfnm[ J.L.]},
  \bibinfo{author}{L\'azaro\xfnm[ F.J.]}.
\newblock \bibinfo{title}{Langevin-dynamics study of the dynamical properties
  of small magnetic particles}.
\newblock \emph{\bibinfo{journal}{Phys Rev B}};
  \bibinfo{year}{1998};\bibinfo{volume}{58}:\bibinfo{pages}{14937--14958}.
\newblock \URLprefix \url{https://link.aps.org/doi/10.1103/PhysRevB.58.14937};
  \DOIprefix\doi{10.1103/PhysRevB.58.14937}.
\bibitem[{Garcia-Palacios(2007)}]{LEcturenotes1}
\bibinfo{author}{Garcia-Palacios\xfnm[ J.L.]}.
\newblock \bibinfo{title}{Introduction to the theory of stochastic processes
  and brownian motion problems}.
\newblock \bibinfo{year}{2007};\URLprefix \url{arXiv:cond-mat/0701242
  [cond-mat.stat-mech]};
  \DOIprefix\doi{https://doi.org/10.48550/arXiv.cond-mat/0701242}.
\bibitem[{Chandrasekhar(1943)}]{RevModPhys.15.1}
\bibinfo{author}{Chandrasekhar\xfnm[ S.]}.
\newblock \bibinfo{title}{Stochastic problems in physics and astronomy}.
\newblock \emph{\bibinfo{journal}{Rev Mod Phys}};
  \bibinfo{year}{1943};\bibinfo{volume}{15}:\bibinfo{pages}{1--89}.
\newblock \URLprefix \url{https://link.aps.org/doi/10.1103/RevModPhys.15.1};
  \DOIprefix\doi{10.1103/RevModPhys.15.1}.
\bibitem[{Kalmykov et~al.(2007)Kalmykov, Coffey and Titov}]{PhysRevE.76.051104}
\bibinfo{author}{Kalmykov\xfnm[ Y.P.]}, \bibinfo{author}{Coffey\xfnm[ W.T.]},
  \bibinfo{author}{Titov\xfnm[ S.V.]}.
\newblock \bibinfo{title}{Phase-space formulation of the nonlinear longitudinal
  relaxation of the magnetization in quantum spin systems}.
\newblock \emph{\bibinfo{journal}{Phys Rev E}};
  \bibinfo{year}{2007};\bibinfo{volume}{76}:\bibinfo{pages}{051104}.
\newblock \URLprefix \url{https://link.aps.org/doi/10.1103/PhysRevE.76.051104};
  \DOIprefix\doi{10.1103/PhysRevE.76.051104}.
\bibitem[{Narducci et~al.(1975)Narducci, Bowden, Bluemel and
  Carrazana}]{PhysRevA.11.280}
\bibinfo{author}{Narducci\xfnm[ L.M.]}, \bibinfo{author}{Bowden\xfnm[ C.M.]},
  \bibinfo{author}{Bluemel\xfnm[ V.]}, \bibinfo{author}{Carrazana\xfnm[ G.P.]}.
\newblock \bibinfo{title}{Phase-space description of the thermal relaxation of
  a ($2j+1$)-level system}.
\newblock \emph{\bibinfo{journal}{Phys Rev A}};
  \bibinfo{year}{1975};\bibinfo{volume}{11}:\bibinfo{pages}{280--287}.
\newblock \URLprefix \url{https://link.aps.org/doi/10.1103/PhysRevA.11.280};
  \DOIprefix\doi{10.1103/PhysRevA.11.280}.
\bibitem[{Brown(1963)}]{PhysRev.130.1677}
\bibinfo{author}{Brown\xfnm[ W.F.]}.
\newblock \bibinfo{title}{Thermal fluctuations of a single-domain particle}.
\newblock \emph{\bibinfo{journal}{Phys Rev}};
  \bibinfo{year}{1963};\bibinfo{volume}{130}:\bibinfo{pages}{1677--1686}.
\newblock \URLprefix \url{https://link.aps.org/doi/10.1103/PhysRev.130.1677};
  \DOIprefix\doi{10.1103/PhysRev.130.1677}.
\bibitem[{Bitaghsir~Fadafan et~al.(2009)Bitaghsir~Fadafan, Liu, Rajagopal and
  Wiedemann}]{BitaghsirFadafan:2008adl}
\bibinfo{author}{Bitaghsir~Fadafan\xfnm[ K.]}, \bibinfo{author}{Liu\xfnm[ H.]},
  \bibinfo{author}{Rajagopal\xfnm[ K.]}, \bibinfo{author}{Wiedemann\xfnm[
  U.A.]}.
\newblock \bibinfo{title}{{Stirring Strongly Coupled Plasma}}.
\newblock \emph{\bibinfo{journal}{Eur Phys J C}};
  \bibinfo{year}{2009};\bibinfo{volume}{61}:\bibinfo{pages}{553--567}.
\newblock \DOIprefix\doi{10.1140/epjc/s10052-009-0885-6};
  \href{http://arxiv.org/abs/0809.2869}{\tt arXiv:0809.2869}.
\bibitem[{Hongo et~al.(2022)Hongo, Huang, Kaminski, Stephanov and
  Yee}]{Hongo:2022izs}
\bibinfo{author}{Hongo\xfnm[ M.]}, \bibinfo{author}{Huang\xfnm[ X.G.]},
  \bibinfo{author}{Kaminski\xfnm[ M.]}, \bibinfo{author}{Stephanov\xfnm[ M.]},
  \bibinfo{author}{Yee\xfnm[ H.U.]}.
\newblock \bibinfo{title}{{Spin relaxation rate for heavy quarks in weakly
  coupled QCD plasma}}.
\newblock \emph{\bibinfo{journal}{JHEP}};
  \bibinfo{year}{2022};\bibinfo{volume}{08}:\bibinfo{pages}{263}.
\newblock \DOIprefix\doi{10.1007/JHEP08(2022)263};
  \href{http://arxiv.org/abs/2201.12390}{\tt arXiv:2201.12390}.
\bibitem[{Hidaka et~al.(2024)Hidaka, Hongo, Stephanov and Yee}]{Hidaka:2023oze}
\bibinfo{author}{Hidaka\xfnm[ Y.]}, \bibinfo{author}{Hongo\xfnm[ M.]},
  \bibinfo{author}{Stephanov\xfnm[ M.A.]}, \bibinfo{author}{Yee\xfnm[ H.U.]}.
\newblock \bibinfo{title}{{Spin relaxation rate for baryons in a thermal pion
  gas}}.
\newblock \emph{\bibinfo{journal}{Phys Rev C}};
  \bibinfo{year}{2024};\bibinfo{volume}{109}(\bibinfo{number}{5}):\bibinfo{pages}{054909}.
\newblock \DOIprefix\doi{10.1103/PhysRevC.109.054909};
  \href{http://arxiv.org/abs/2312.08266}{\tt arXiv:2312.08266}.
\bibitem[{Chun et~al.(2021)Chun, Gao and Horowitz}]{PhysRevResearch.3.043172}
\bibinfo{author}{Chun\xfnm[ H.M.]}, \bibinfo{author}{Gao\xfnm[ Q.]},
  \bibinfo{author}{Horowitz\xfnm[ J.M.]}.
\newblock \bibinfo{title}{Nonequilibrium green-kubo relations for hydrodynamic
  transport from an equilibrium-like fluctuation-response equality}.
\newblock \emph{\bibinfo{journal}{Phys Rev Res}};
  \bibinfo{year}{2021};\bibinfo{volume}{3}:\bibinfo{pages}{043172}.
\newblock \URLprefix
  \url{https://link.aps.org/doi/10.1103/PhysRevResearch.3.043172};
  \DOIprefix\doi{10.1103/PhysRevResearch.3.043172}.
\bibitem[{Daniel(1997)}]{Herbert:1997}
\bibinfo{author}{Daniel\xfnm[ H.]}.
\newblock \bibinfo{title}{Physik: Elektrodynamik - relativistische Physik};
  vol.~\bibinfo{volume}{2}.
\newblock \bibinfo{publisher}{Walter de Gruyter}; \bibinfo{year}{1997}.
\bibitem[{Tuchin(2013{\natexlab{a}})}]{Tuchin:2013ie}
\bibinfo{author}{Tuchin\xfnm[ K.]}.
\newblock \bibinfo{title}{{Particle production in strong electromagnetic fields
  in relativistic heavy-ion collisions}}.
\newblock \emph{\bibinfo{journal}{Adv High Energy Phys}};
  \bibinfo{year}{2013}{\natexlab{a}};\bibinfo{volume}{2013}:\bibinfo{pages}{490495}.
\newblock \DOIprefix\doi{10.1155/2013/490495};
  \href{http://arxiv.org/abs/1301.0099}{\tt arXiv:1301.0099}.
\bibitem[{Huang et~al.(2023)Huang, She, Shi, Huang and Liao}]{Huang:2022qdn}
\bibinfo{author}{Huang\xfnm[ A.]}, \bibinfo{author}{She\xfnm[ D.]},
  \bibinfo{author}{Shi\xfnm[ S.]}, \bibinfo{author}{Huang\xfnm[ M.]},
  \bibinfo{author}{Liao\xfnm[ J.]}.
\newblock \bibinfo{title}{{Dynamical magnetic fields in heavy-ion collisions}}.
\newblock \emph{\bibinfo{journal}{Phys Rev C}};
  \bibinfo{year}{2023};\bibinfo{volume}{107}(\bibinfo{number}{3}):\bibinfo{pages}{034901}.
\newblock \DOIprefix\doi{10.1103/PhysRevC.107.034901};
  \href{http://arxiv.org/abs/2212.08579}{\tt arXiv:2212.08579}.
\bibitem[{Jiang et~al.(2022)Jiang, Cao, Xing, Wu, Yang and
  Zhang}]{PhysRevC.105.054907}
\bibinfo{author}{Jiang\xfnm[ Z.F.]}, \bibinfo{author}{Cao\xfnm[ S.]},
  \bibinfo{author}{Xing\xfnm[ W.J.]}, \bibinfo{author}{Wu\xfnm[ X.Y.]},
  \bibinfo{author}{Yang\xfnm[ C.B.]}, \bibinfo{author}{Zhang\xfnm[ B.W.]}.
\newblock \bibinfo{title}{Probing the initial longitudinal density profile and
  electromagnetic field in ultrarelativistic heavy-ion collisions with heavy
  quarks}.
\newblock \emph{\bibinfo{journal}{Phys Rev C}};
  \bibinfo{year}{2022};\bibinfo{volume}{105}:\bibinfo{pages}{054907}.
\newblock \URLprefix
  \url{https://link.aps.org/doi/10.1103/PhysRevC.105.054907};
  \DOIprefix\doi{10.1103/PhysRevC.105.054907}.
\bibitem[{Huang et~al.(2018)Huang, Jiang, Shi, Liao and Zhuang}]{Huang:2017tsq}
\bibinfo{author}{Huang\xfnm[ A.]}, \bibinfo{author}{Jiang\xfnm[ Y.]},
  \bibinfo{author}{Shi\xfnm[ S.]}, \bibinfo{author}{Liao\xfnm[ J.]},
  \bibinfo{author}{Zhuang\xfnm[ P.]}.
\newblock \bibinfo{title}{{Out-of-equilibrium chiral magnetic effect from
  chiral kinetic theory}}.
\newblock \emph{\bibinfo{journal}{Phys Lett B}};
  \bibinfo{year}{2018};\bibinfo{volume}{777}:\bibinfo{pages}{177--183}.
\newblock \DOIprefix\doi{10.1016/j.physletb.2017.12.025};
  \href{http://arxiv.org/abs/1703.08856}{\tt arXiv:1703.08856}.
\bibitem[{McLerran and Skokov(2014)}]{McLerran:2013hla}
\bibinfo{author}{McLerran\xfnm[ L.]}, \bibinfo{author}{Skokov\xfnm[ V.]}.
\newblock \bibinfo{title}{{Comments About the Electromagnetic Field in
  Heavy-Ion Collisions}}.
\newblock \emph{\bibinfo{journal}{Nucl Phys A}};
  \bibinfo{year}{2014};\bibinfo{volume}{929}:\bibinfo{pages}{184--190}.
\newblock \DOIprefix\doi{10.1016/j.nuclphysa.2014.05.008};
  \href{http://arxiv.org/abs/1305.0774}{\tt arXiv:1305.0774}.
\bibitem[{Kapusta et~al.(2020)Kapusta, Rrapaj and Rudaz}]{Kapusta:2019sad}
\bibinfo{author}{Kapusta\xfnm[ J.I.]}, \bibinfo{author}{Rrapaj\xfnm[ E.]},
  \bibinfo{author}{Rudaz\xfnm[ S.]}.
\newblock \bibinfo{title}{{Relaxation Time for Strange Quark Spin in Rotating
  Quark-Gluon Plasma}}.
\newblock \emph{\bibinfo{journal}{Phys Rev C}};
  \bibinfo{year}{2020};\bibinfo{volume}{101}(\bibinfo{number}{2}):\bibinfo{pages}{024907}.
\newblock \DOIprefix\doi{10.1103/PhysRevC.101.024907};
  \href{http://arxiv.org/abs/1907.10750}{\tt arXiv:1907.10750}.
\bibitem[{Acharya et~al.(2021)}]{ALICE:2020iev}
\bibinfo{author}{Acharya\xfnm[ S.]}, et~al. (\bibinfo{collaboration}{ALICE}).
\newblock \bibinfo{title}{{First measurement of quarkonium polarization in
  nuclear collisions at the LHC}}.
\newblock \emph{\bibinfo{journal}{Phys Lett B}};
  \bibinfo{year}{2021};\bibinfo{volume}{815}:\bibinfo{pages}{136146}.
\newblock \DOIprefix\doi{10.1016/j.physletb.2021.136146};
  \href{http://arxiv.org/abs/2005.11128}{\tt arXiv:2005.11128}.
\bibitem[{Weickgenannt and Blaizot(2025)}]{Weickgenannt:2024ibf}
\bibinfo{author}{Weickgenannt\xfnm[ N.]}, \bibinfo{author}{Blaizot\xfnm[
  J.P.]}.
\newblock \bibinfo{title}{{Spin kinetic theory with a nonlocal relaxation time
  approximation}}.
\newblock \emph{\bibinfo{journal}{Phys Rev D}};
  \bibinfo{year}{2025};\bibinfo{volume}{111}(\bibinfo{number}{5}):\bibinfo{pages}{056006}.
\newblock \DOIprefix\doi{10.1103/PhysRevD.111.056006};
  \href{http://arxiv.org/abs/2409.11045}{\tt arXiv:2409.11045}.
\bibitem[{Gursoy et~al.(2014)Gursoy, Kharzeev and Rajagopal}]{Gursoy:2014aka}
\bibinfo{author}{Gursoy\xfnm[ U.]}, \bibinfo{author}{Kharzeev\xfnm[ D.]},
  \bibinfo{author}{Rajagopal\xfnm[ K.]}.
\newblock \bibinfo{title}{{Magnetohydrodynamics, charged currents and directed
  flow in heavy ion collisions}}.
\newblock \emph{\bibinfo{journal}{Phys Rev C}};
  \bibinfo{year}{2014};\bibinfo{volume}{89}(\bibinfo{number}{5}):\bibinfo{pages}{054905}.
\newblock \DOIprefix\doi{10.1103/PhysRevC.89.054905};
  \href{http://arxiv.org/abs/1401.3805}{\tt arXiv:1401.3805}.
\bibitem[{Tuchin(2013{\natexlab{b}})}]{Tuchin:2012mf}
\bibinfo{author}{Tuchin\xfnm[ K.]}.
\newblock \bibinfo{title}{{Electromagnetic radiation by quark-gluon plasma in a
  magnetic field}}.
\newblock \emph{\bibinfo{journal}{Phys Rev C}};
  \bibinfo{year}{2013}{\natexlab{b}};\bibinfo{volume}{87}(\bibinfo{number}{2}):\bibinfo{pages}{024912}.
\newblock \DOIprefix\doi{10.1103/PhysRevC.87.024912};
  \href{http://arxiv.org/abs/1206.0485}{\tt arXiv:1206.0485}.

\end{thebibliography}
\end{document}